\documentclass[10pt,notitlepage,nobibnotes,nofootinbib]{revtex4-1}
\usepackage[latin1]{inputenc}
\usepackage{amsmath}
\usepackage{amsfonts}
\usepackage{amssymb}
\usepackage{graphicx}
\usepackage{bm}
\usepackage{float}
\usepackage{subfig}
\usepackage{tikz}
\usetikzlibrary{calc}

\usepackage[colorlinks=true,citecolor=blue,urlcolor=blue]{hyperref}
\usepackage[capitalize]{cleveref} 

\newcommand{\TTV}{\bm{\theta}}
\newcommand{\TT}{\theta}

\renewcommand{\Re}{\operatorname{Re}}
\renewcommand{\Im}{\operatorname{Im}}

\newcommand{\EE}{\mathbb{E}}
\renewcommand{\SS}{L}
\newcommand{\RR}{{\L}}
\newcommand{\QQ}{\mathfrak{H}}
\newcommand{\qq}{\mathfrak{h}}
\newcommand{\QQR}{\mathfrak{R}}
\newcommand{\ff}{\mathfrak{f}}
\newcommand{\FF}{\mathfrak{F}}
\renewcommand{\TH}{\hat{{\theta}}}
\newcommand{\THV}{\bm{\hat{\theta}}}
\newcommand{\tr}{\mathrm{Tr}}
\newcommand{\var}{\mathrm{Var}}

  \newcommand{\II}{\mathbb{I}}

 \newcommand{\bra}[1]{\langle #1 \vert}
 \newcommand{\ket}[1]{\vert #1 \rangle}

 \newcommand{\twobytwo}[4]{\left( \begin{smallmatrix} #1&#2\\ #3&#4 \end{smallmatrix} \right)}
 \newcommand{\threebythree}[9]{\left( \begin{smallmatrix} #1&#2&#3\\#4&#5&#6\\#7&#8&#9 \end{smallmatrix} \right)}

\begin{document}
	
		\title{Uncertainty and Trade-offs in Quantum Multiparameter Estimation}
	
	\author{Ilya Kull} 
	\affiliation{Faculty of Physics, University of Vienna, Boltzmanngasse 5, 1090 Vienna, Austria}
	\affiliation{Institute for Quantum Optics and Quantum Information (IQOQI),	Austrian Academy of Sciences, Boltzmanngasse 3, 1090 Vienna, Austria}	
	
	 	\author{Philippe Allard Gu\'erin} 
	 \affiliation{Faculty of Physics, University of Vienna, Boltzmanngasse 5, 1090 Vienna, Austria}
	 \affiliation{Institute for Quantum Optics and Quantum Information (IQOQI),	Austrian Academy of Sciences, Boltzmanngasse 3, 1090 Vienna, Austria} 
	\author{Frank Verstraete} 
	 \affiliation{Department of Physics and Astronomy, Ghent University, Krijgslaan 281, 9000 Gent, Belgium}

	\begin{abstract}
		Uncertainty relations in quantum mechanics express bounds on our ability to simultaneously obtain knowledge about expectation values of non-commuting observables of a quantum system. They quantify  trade-offs in    accuracy between complementary pieces of information about the system. 		
		In Quantum multiparameter estimation, 	such trade-offs occur for the precision achievable for different parameters characterizing a density matrix: an uncertainty relation emerges between the achievable variances of the different estimators. This is in contrast to classical multiparameter estimation, where simultaneous optimal precision is attainable in the asymptotic limit. 
		We study trade-off relations that follow from  known tight  bounds in quantum multiparameter estimation.  We compute trade-off curves and  surfaces from  Cram\' er--Rao type bounds which provide a compelling  graphical representation of the information encoded in such bounds, and argue that bounds on simultaneously achievable precision in quantum multiparameter estimation   should be regarded  as  measurement uncertainty relations.  
	  {From the  state-dependent bounds on the expected cost in  parameter estimation, we derive a  \textit{state independent}   uncertainty relation   between the parameters of a qubit system.}
	\end{abstract}
	
	\maketitle
	
	Ever since its first formulation, the uncertainty principle has seen many refinements and clarifications. 
	As quantum theory developed, its state-of-the-art concepts and mathematical tools were used to  formulate in precise terms the ideas which were put forward in Heisenberg's 1927 paper~\cite{Heisenberg1927}. 	
	As a result, our current understanding of the  uncertainties inherent in quantum mechanics is spelled out  in a collection of theorems pertaining to well defined operational tasks.

	Soon after Heisenberg's paper, rigorous proofs of his uncertainty relations were formulated~\cite{Kennard1927,Weyl1928,Robertson1929}. Those are usually referred to as \textit{preparation uncertainty relations}. Most well known  is the    relation due to Weyl  and Robertson 
\begin{equation} \label{eq:Robertson}
		{\sigma_A}{\sigma_B}\geq \frac{1}{2} | \left\langle  [A,B] \right\rangle | \ ,
\end{equation}
		 where 
		$\sigma_A= \sqrt{\left\langle  (A-\langle A\rangle)^2 \right\rangle}$ is the standard deviation of an observable $A$ in a given state $\psi$ ($\langle \cdot\rangle\equiv\bra{\psi}\cdot\ket{\psi}$).
		For canonically conjugate observables such as position and momentum the right hand side of~\cref{eq:Robertson} equals $\hbar/2$. This relation implies that  
		it is impossible  \emph{to prepare} a particle in a state with arbitrarily sharp statistics for both position and momentum.
		Note that such uncertainty relations do not tell anything about statistics of joint measurements of both observables. Rather, the standard deviations on the left hand side of~\cref{eq:Robertson} correspond to measurements of $A$ and $B$ on two independent ensembles of identical copies of the state $\ket{\psi}$. The preparation uncertainty relation between position and momentum is tight, as equality is achieved for specific states \cite{BUSCH2007155}. The relation~\cref{eq:Robertson} hence quantifies an attainable  trade off   between the sharpness of the position and  momentum measurement statistics.  
		Subsequent works formulated preparation uncertainty relations which involve other measures for the spread of a distribution \cite{Maassen88,Landau61,Uffink1985,Coles2017}. 

		The development of quantum measurement theory~\cite{Kraus1974,Davies1976,Kraus1983} allowed 
		to formulate \textit{accuracy--disturbance} uncertainty relations which quantify the disturbance caused by a positive operator valued measure (POVM) measurement   
				 to the statistics of a subsequent measurement of another POVM~\cite{Martens_1992,PhysRevA.53.2038,OZAWA200321,OZAWA2004350,Ozawa_2005,Hashagen2019,Renes2017uncertainty}. 
		\textit{Joint measurement uncertainty relations} have been discussed by many authors~\cite{ISHIKAWA1991257,doi:10.1119/1.17657,Appleby:1998ce,PhysRevD.33.2253,1985IJTP...24...63B} and most recently in Ref.~\cite{BuschLahtiWerner}. They describe the deviation of the statistics in a  joint approximate  measurement of two quantities from their statistics when  measured separately.
	Many more authors have considered  these two notions of uncertainty, for a more complete list see references in Refs.~\cite{Hashagen2019,Renes2017uncertainty}. There is still debate between the proponents of the most recent approaches regarding  which of them  most properly  captures Heisenberg's qualitative considerations~\cite{PhysRevLett.111.160405, BuschLahtiWernerRMS, sep-qt-uncertainty,2013arXiv1308.3540O,2016Entrp..18..174A}.

	Quantum parameter estimation theory  provides yet another way to quantify quantum uncertainty. In this framework one considers a family of quantum states parametrized by real numbers, and the task is to estimate the parameters corresponding to a given state  by performing measurements on identical copies of the state. 
	In the one parameter case, the quantum Fisher information (QFI) Cram\' er--Rao bound    provides a lower bound on the asymptotic scaling of the variance of an unbiased estimator~\cite{Helstrom,Holevo1982}. The bound is  achievable in the asymptotic limit of many copies of the state with a separable measurement~\cite{BraunsteinCaves,Barndorff_Nielsen_2000}. 
	Of particular importance is the case when the parameter to be estimated is elapsed time $t$ for a state $|\psi(t)\rangle=\exp(-itH)|\psi(0)\rangle$ evolved with a given Hamiltonian; in that case, the quantum Cram\' er--Rao bound is proportional to the expectation value of the Hamiltonian~\cite{BRAUNSTEIN1996135}, and hence yields the well known time-energy uncertainty relation~\cite{MandelstamTamm}.
	Results  of this type can be seen as \textit{hybrid preparation--measurement uncertainty relations}, as they describe a trade off between the accuracy of a \textit{measured} quantity, namely, the     estimator for the desired parameter;  and  the variance of the   operator   generating translations in that parameter, a quantity  pertaining  to  the preparation. 
	Quantum parameter  estimation has been also used to formulate joint measurement uncertainty relations \cite{Watanabe2011,Zhu2015} and  error-disturbance relations \cite{Shitara2016}.


	Classically,  going from  single parameter estimation to a multiparameter setting  involves replacing the scalar Cram\' er--Rao bound by  a matrix inequality.  
	 This multiparameter bound is still asymptotically achievable~\cite{Cramer}, which means that the optimal precision can be achieved for all parameters \textit{simultaneously}. 
	 In  quantum multiparameter estimation however, the quantum Cram\' er--Rao bound is in general no longer attainable as the measurements required to attain the single parameter  bound for the individual parameters might not be compatible~\cite{Ragy}. In this setting one expects there to be   \textit{trade-offs} between the precision achievable for the estimators of  different parameters. This is clearly a pure quantum phenomenon, and such trade-offs should hence be viewed as yet another manifestation of quantum uncertainty. Such bounds on quantum multiparameter estimation belong to  the    measurement type of uncertainty relations.
	 The `no go' part of such uncertainty relations  is  the unattainability of the   multiparameter QFI Cram\' er--Rao bound. It implies that, in contrast to  the classical case,   optimal precision for all parameters simultaneously is impossible to achieve---acquiring better statistics for one parameter automatically leads to worse statistics for the complementary ones.  The positive content is the characterization of the achievable trade-off and the measurement schemes attaining it. 
	 Various   bounds on quantum multiparameter estimation that appear in the literature already encode such trade-off  relations. The aim of this paper is to focus attention on this particular feature of the known tight bounds.
	 
	A point of distinction from   other kinds of uncertainty relations is that in quantum multiparameter estimation, the quantities that one tries to estimate do have simultaneously well defined values. The task is to uncover classical information encoded in the state, e.g.\ the settings that were chosen on  the device that prepared the state. Furthermore, given arbitrarily many copies of the state, all of the parameters can be estimated with arbitrary precision. Trade-offs appear when one considers the precision  increase for each unknown parameter \textit{per copy} of the state.  A sharp distinction  has to be also made between bounds for separable measurements---the realistic situation in experiments---versus collective measurements---which  involve  highly entangled measurements between the different copies.

		Quantum multiparameter  estimation has been an active field of research for nearly five decades.  It has seen significant  recent   development which was stimulated in part by the increasing  relevance of multiparameter estimation to quantum metrology tasks. We do not attempt to provide a comprehensive   review of the field. Rather, we present the minimal background  needed for the presentation of our results in a self contained manner. For a proper and up-to-date introduction to the field we refer to several very recent reviews which cover the state-of-the-art  theoretical results as well as   applications to concrete tasks \cite{Liu_2019,Albarelli2019,sidhu2019geometric,demkowiczdobrzanski2020multiparameter}.

	Attainable bounds for multiparameter estimation are known  for several quantum statistical models. For estimation of shift parameters of  Gaussian states,   Holevo proved an  achievable bound~\cite{Holevo1982}; this bound is referred to in the literature as the Holevo Cram\' er--Rao  bound~\cite{Ragy,HayashiMatsumoto}. The theory of local asymptotic normality~\cite{2009CMaPh.289..597K,yamagata2013} implies that this bound is  achievable  for finite dimensional quantum systems if one allows \textit{collective} measurements to be performed on many identical  copies of the state.
		  Attainable bounds for a qubit system have been proven by Nagaoka, Hayashi, and  Gill and Massar~\cite{Nagaoka,Hayashi,GillMassar}. 
	  Attainability of the quantum Fisher information Cram\'er--Rao bound with collective measurements has been shown to be equivalent to what is called the commutation condition~\cite{Ragy}, which involves the commutators of the operators whose measurement provides the optimal one parameter precision.

	In this paper, we translate the various known bounds on quantum multiparameter estimation into trade-off curves (or hyper-surfaces, in the case of more that two parameters). Such curves provide a visually clear representation of the information encoded in  bounds on estimation. They highlight the trade-off, which is not evident when the bounds  are written down   as inequalities. 
	To demonstrate this, we use trade-off curves   to compare the bounds for estimation in a qubit system such as the  Gill--Massar bound (which is attainable with separable measurements) and the Holevo Cram\' er--Rao bound (which is only attainable with collective measurements). 
		We  show how to sample points from the trade-off surface corresponding to the Gill--Massar bound for different parametrizations of a qubit state and discuss the family of measurements which attain the bound in different parametrizations.

		Our main result is the derivation of a   \textit{state independent}  trade-off relation between the three parameters of a qubit system when estimated using separable measurements. This result follows from the Gill--Massar bound which is state-dependent---like many other Cram\' er--Rao type bounds in quantum parameter estimation.
	Our state independent  result is obtained by superimposing the trade-off surfaces corresponding to different states in the same plot to obtain a region in the $3$-dimensional space of the variances of the estimators   which is unattainable for all states and all separable measurements. 
	This result implies a \textit{state independent measurement uncertainty relation} between the  three Pauli operators $\sigma_x,\sigma_y,\sigma_z$. We prove the corresponding uncertainty relation for two parameters which turns out to have a simple additive form $\var(\TH_i)+\var(\TH_j)\geq1/4N$, for  $i\neq j\in\{x,y,z\}$ and where $\var(\TH_i)$ is the variance of the estimator for the parameter $\TT_i=\langle \sigma_i/2\rangle$ and $N$ is the number of copies of the state. We further show that the bound $\var(\TH_x)+\var(\TH_y)+\var(\TH_z)\geq1/ N$ holds and forms part of the trade-off surface.

		 Finally, we compute the Holevo Cram\' er--Rao bound for a three level system model and describe the structure of its trade-off surface which is generic to any $d$-dimensional quantum system.

	The paper is organized as follows. In~\cref{sec:perlim} we briefly review the required background and set up our notation; in~\cref{sec:TradeOff} we show how to obtain trade-off curves from Cram\' er--Rao type bounds for two parameters; in~\cref{sec:2levelSys} we present our result for the qubit model which include a state independent trade-off surface; \cref{sec:3levelSys}  describes the structure of the trade-off surface of the Holevo Cram\' er--Rao bound for a qutrit; we conclude with a discussion in~\cref{sec:Discussion}.

	\section{Preliminaries}
	\label{sec:perlim}
	We start by reviewing  estimation theory. In classical estimation theory~\cite{Cramer} we are given a family of probability distributions with probability density $p(\TTV)$ parametrized by a vector of parameters  $\TTV=(\TT_1,\ldots,\TT_K)$. The task is to estimate  the unknown values   $\TTV_0$ by sampling from $p(\TTV_0)$.
	In order to do so we shall pick an estimator,  a function that produces an estimated value $\THV(x_1,x_2,\ldots,x_N)$ given the $N$ samples drawn $\{x_i\}$. The estimation statistics are  then described by the random variable $\THV(X_1,X_2,\ldots,X_N)$, where the random variables $X_i$ are distributed according to $p(x_i|\TTV_0):=p(X=x_i|\TTV_0)$. An estimator is called locally unbiased if $\EE \THV=\TTV_0$, where $\EE$ is the expectation value is with respect to $p(\TTV_0)$.

\subsection{	The Cram\' er--Rao  Bound}
 Let $\TT$ be a single parameter. The Cram\' er--Rao bound  is a lower bound on the variance of the estimator $\var(\TH)=\EE(\TH-\TT_0)^2$. When the estimator is unbiased the bound is given by the inverse of the Fisher information  
	$\ff(\TT_0) := \left.\EE \left( \frac{d\log p(x|\TT)}{d\TT}\right)^2\right|_{\TT=\TT_0}$
	\begin{equation} \label{eq:1paramCRBound}
	\var(\TH)\geq \frac{1}{N\ff(\TT_0) }  \ ,
	\end{equation}
with $N$ the number of samples.
	 
	In the multiparameter case we define the covariance matrix of the estimators (we shall suppress the $\TT_0$ dependence in the notation)
	 \[  V(\THV)_{ij}=\EE (\TH_i-\TT_i)(\TH_j-\TT_j) \ ,
	 \] 
	 and  the Fisher information matrix
	 \begin{equation} \label{eq:FisherInfoDef}
	   \FF_{ij} = \EE \frac{d\log p}{d\TT_i}\frac{d\log p}{d\TT_j} \ .
	 \end{equation}
	 The Cram\' er--Rao bound then takes the form of an inequality between positive semidefinite matrices
	 \begin{equation} \label{eq:MultiParamCRBound}
	 V(\THV)\geq\FF(\TTV_0)^{-1} /N\ .
	 \end{equation}
 	This bound is achievable asymptotically by the maximum likelihood method. More precisely, it is shown that there is a locally unbiased estimator for which the rescaled covariance matrix  $NV \approx \FF^{-1}$ in the limit of large $N$~\cite{Cramer}.  
	To compensate for the overall $1/N$ improvement in precision due to the use of many copies of the source $p(\TTV)$,
	we pick the  rescaled covariance matrix $NV$ as the figure of merit for the precision of the estimator in the asymptotic regime. We keep the $N$ explicit in the notation as a reminder.	
 
	\subsection{Quantum Parameter Estimation} 
	\label{sec:perlimB}
	 In quantum parameter estimation,  instead of a probability distribution we are given a quantum state $\rho(\TT)$ (satisfying $\rho\geq 0, \ \tr\rho = 1$) which depends on $\TT$. For a given measurement $\bm{M}$ with POVM elements $\{M_i\}$ (satisfying $M_i\geq 0, \  \sum M_i =\II$) we obtain a probability distribution for the outcomes $p^{\bm{M}}(i|\TT)=\tr M_i\rho(\TT)$ which depends on $\TT$ through  the state $\rho(\TT)$. Classical estimation  theory can now be applied to the estimation of  $\TT$ from $p^{\bm{M}}$. The problem of quantum parameter estimation is hence equivalent to the one of finding the measurement which maximizes this classical Fisher information.  The Fisher information associated with the measurement $\bm{M}$ is 
	\begin{equation} \label{eq:QFisherInfoDef}
		\ff^{\bm{M}}(\TT_0) := \EE \left(\frac{d\log p^{\bm{M}}}{d\TT}\right)^2 =
		 \sum_i \frac{(\tr M_i \frac{d \rho}{d\TT} )^2}{\tr M_i\rho(\TT_0)} \ ,
	\end{equation}
	where $\frac{d \rho}{d\TT}$ is evaluated at $\TT_0$.
	The  symmetric logarithmic derivative quantum Fisher information (SLD-QFI) is defined as
	 \begin{equation*}
	 \qq(\TT_0) = \tr (\rho(\TT_0) \SS(\TT_0)^2) \ ,
	 \end{equation*}
	  where $\SS(\TT_0)$ is the symmetric logarithmic derivative  (SLD) defined implicitly by
	\begin{equation} \label{eq:SLDdef}
 \left.	\frac{d \rho}{d\TT}\right|_{\TT_0} =:\frac{1}{2} (\SS(\TT_0)\rho(\TT_0) + \rho (\TT_0)\SS(\TT_0)) \ .
	\end{equation}
	When $\rho$ is of full rank, solutions to~\cref{eq:SLDdef} are unique as the only matrix that anti-commutes with $\rho$ is the zero matrix. We will always assume that this is the case. For treatment of the case of degenerate states see Refs.~\cite{BraunsteinCaves,Matsumoto_2002,ERCOLESSI20131996}.
	  The SLD-QFI bound~\cite{BraunsteinCaves} states that for any measurement $\bm{M}$  
	\begin{equation} \label{eq:OneParamQFI}
	 	\ff^{\bm{M}}(\TT_0)\leq \qq (\TT_0) \ ,
	\end{equation}
	 (we  shall suppress the $\TT_0$ dependence from now on).
	 The proof is obtained by the use of the Cauchy--Schwarz inequality:
	 \begin{equation*}
	 \begin{split}
	 \ff^{\bm{M}} =& \sum_i \frac{(\tr M_i \frac{d \rho}{d\TT} )^2}{\tr M_i\rho} = 
	 				 \sum_i \frac{\Re \left(\tr M_i  \rho\SS \right)^2}{\tr M_i\rho} \leq
	 				 \sum_i \frac{\left| \tr M_i  \rho\SS \right|^2}{\tr M_i\rho} = \\
	 			& 	  \sum_i \frac{\left| \tr \sqrt{M_i}\sqrt{M_i}  \sqrt{\rho}\sqrt{\rho}\SS \right|^2}{\tr M_i\rho} \leq
	 			 	  \sum_i \frac{ \tr( M_i \rho) \tr( \rho\SS M_i\ \SS) }{\tr M_i\rho}   =  \\
	 			&  \tr (\rho \SS^2)  = \qq \ , 
	 \end{split}
	 \end{equation*}
	 where we used the definition of the SLD in the second equality and  $\sum M_i =\II$ in going to the last line. 
	Braunstein and Caves~\cite{BraunsteinCaves} proved that equality in~\cref{eq:OneParamQFI} is attained when $\bm{M}$ is a projective measurement in the basis which diagonalizes $\SS$, hence identifying the optimal measurement strategy. 
	
	One can also define  the right logarithmic derivative (RLD) and corresponding to it is the right logarithmic derivative  quantum Fisher information (RLD-QFI) bound. This bound will be discussed later.

	In the case of multiple parameters, the Fisher information matrix of the measurement $\bm{M}$  is defined according to~\cref{eq:FisherInfoDef}  as
		\begin{equation*} 
	\FF^{\bm{M}}_{ij} := \EE \frac{d\log p^{\bm{M}}}{d\TT_i}\frac{d\log p^{\bm{M}}}{d\TT_j} =
	\sum_k \frac{\tr M_k \frac{d \rho}{d\TT_i} \tr M_k \frac{d \rho}{d\TT_j} }{\tr M_k\rho}  \ .
	\end{equation*}
	The quantum Fisher information matrix is defined as
	\begin{equation*}
		\QQ_{ij} = \frac{1}{2}\tr \rho (\SS_i\SS_j+\SS_i\SS_j) \ , 
	\end{equation*}
	where $\SS_i$ is the symmetric logarithmic derivative with respect to $\TT_i$. 
	The multiparameter SLD-QFI bound is an inequality in the sense of semidefinite matrices:
	\begin{equation} \label{eq:MultiParamQFI}
	\FF^{\bm{M}}\leq \QQ  \ .
	\end{equation}
	This bound is a consequence of the one parameter bound~\cref{eq:OneParamQFI}. To see this, let $\bm{v}$ be a vector in the space of parameters $\mathbb{R}^K$\footnote
	{
	A note on notation: to reduce confusion between state vectors in Hilbert space and vectors in parameter space we will stick to Dirac notation $\bra{\psi}O \ket{\phi}$ for the former and vector notation $\bm{v}^\intercal M\bm{v}$ for the latter. 	
	},
let   $\TT_{\bm{v}}:=\sum_i v_i \TT_i$.  From linearity of the definition of the SLD~\cref{eq:SLDdef}, it follows that the corresponding symmetric logarithmic derivative is
\begin{equation} \label{eq:LinCombSLD}
\SS_{\bm{v}} = \sum_i v_i \SS_i \ .
\end{equation}
We then have
	\begin{equation*}
	\begin{split}
		\bm{v}^\intercal \FF^{\bm{M}} \bm{v}& =
		\sum_{ijk} {v}_i 
		\frac{\tr M_k \frac{d \rho}{d\TT_i} \tr M_k \frac{d \rho}{d\TT_j} }{\tr M_k\rho} 
		 {v}_j
		 = 
		 \sum_k \frac{\tr M_k \frac{d \rho}{d\TT_{\bm{v}} } \tr M_k \frac{d \rho}{d\TT_{\bm{v}}} }{\tr M_k\rho} 
		 = \ff_{\bm{v}}
		 \leq 
		  \\ 
		  &\qq_{\bm{v}} =	 \tr \rho \SS_{\bm{v}}\SS_{\bm{v}}
		 = 
		\sum_{ij}	{v}_i \frac{1}{2}\tr \rho (\SS_i\SS_j+\SS_i\SS_j) {v}_j 
		= 
		  \bm{v}^\intercal\QQ \bm{v} \ ,
	\end{split}
	\end{equation*}
	where $= \ff^{\bm{M}}_{\bm{v}},$ and $	\qq_{\bm{v}} $ denote the one parameter Fisher information   of the measurement $\bm{M}$  and the  quantum Fisher information    for the estimation of $\TT_{\bm{v}}$ respectively.
	
	In other words, in the multi parameter setting, the SLD-QFI
	 bound~\cref{eq:MultiParamQFI}   can be stated as the following: for any linear combination of the parameters $\TT_{\bm{v}}=\sum v_i\TT_i$, a  one parameter SLD-QFI bound $\ff_{\bm{v}} \leq \qq_{\bm{v}}$  applies. In addition, for any $\bm{v}$  the bound is attainable with a projective measurement in the basis diagonalizing $\SS_{\bm{v}}$\footnote
	{
		Note that for the  measurement $\tilde{M}$ which is optimal for the estimation of the parameter $\TT_1$ in a one parameter setting (i.e.\ when all other parameters are kept fixed)   we have asymptotically $NV\approx{\FF^{\tilde{M}}}^{-1} $ which implies
		\[
		NV_{1} = ({\FF^{\tilde{M}}}^{-1})_{11} \geq (\QQ^{-1})_{11}  \geq 1/\QQ_{11} = 1/\qq_1  = 1/\ff^{\tilde{M}}_1 \ , 
		\]
		where the second inequality is a general property of positive matrices, and the last equality is due to the optimality assumption about $\tilde{M}$.
		That is, the optimal measurement for one parameter when estimated alone might perform worse for the estimation of the same parameter when additional parameters are unknown~\cite{Ragy}.   
	}.

	Further notice that because of their quadratic forms, the covariance matrix $V$, the Fisher information matrix $\FF$, and the quantum Fisher information matrix $\QQ$ all transform in the same way under  linear coordinate transformations. When $\TT_i \mapsto \tilde{\TT_i}:= \sum_jR_{ij}\TT_j$, all three matrices transform as $(\cdot)\mapsto R(\cdot) R^\intercal$. This implies that matrix inequalities between them are invariant under rotations of coordinates\footnote
	{
	Because we are dealing with local estimation, only linear coordinate transformations are of interest (see Ref.~\cite{GillMassar}). An arbitrary (smooth) coordinate transformation  will be approximated to first order by a linear one $\TT_i \mapsto \tilde{\TT_i}(\TTV)= \tilde{\TT_i}(\TTV_0)+ \sum_j\partial_{\TT_j} \tilde{\TT_i}(\TTV_0) (\TT_j-{\TT_0}_j) + o(|\TTV-\TTV_0|)$. 
	}.
		
		\section{Trade-off}
		\label{sec:TradeOff}
	If two linear combinations of the parameters $\{\TT_i\}$ defined by  the vectors $\bm{u}$ and $\bm{v}$  result in commuting SLDs $[\SS_{\bm{u}},\SS_{\bm{v}}]=0$, then optimal estimation of the two parameters $\TT_{\bm{u}}$ and $\TT_{\bm{v}}$ can be achieved simultaneously by performing a measurement in the basis which diagonalizes both of them.
	
	However, $[\SS_{\bm{u}},\SS_{\bm{v}}]=0$ will typically not be satisfied.
	In general, we expect there to be a trade-off between the achievable precision in the two parameters in the following sense. Let $M(\lambda), \ \lambda\in[0,1]$ be a family of measurements with POVM  elements $\{M_i(\lambda)\}$ such that $M(0)$ is the optimal measurement for $\TT_{\bm{v}}$ and $M(1)$ is the optimal measurement for $\TT_{\bm{u}}$. 
	 For intermediate values of  $\lambda$ the precisions of the estimators for $\TT_{\bm{x}}$ (which we quantify by $\var(\TT_{\bm{x}})$) will take values larger than optimal. 
	
	Trade-off curves are commonly used in detection theory. In particular receiver operating characteristic curves (ROC curves) are a convenient way to represent how the probability for false positive detection increases as one increases the sensitivity~\cite{1057460}. In the context of uncertainty relations, a similar representation was used in~\cite{Dammeier_2015} for preparation uncertainties of angular momentum components.   
	As we will now show, trade-off curves (or surfaces) are a convenient representation of the data which is typically encoded in uncertainty relations.
	
	The known bounds on precision in parameter estimation are most often stated as lower bounds on the expected cost, resulting from a given positive definite $K\times K$ cost matrix $G$~\cite{Holevo1982,HayashiMatsumoto,GillMassar}\footnote
	{Once again, because we are dealing with local estimation representing the cost function by a positive  matrix is general enough. Expanding an arbitrary cost function $f(\THV-\TTV_0)$ around the minimum $\TTV_0$ and taking the expectation value we obtain  $\mathbb{E}f(\THV-\TTV_0)=f(\TTV_0)  +
		1/2 \mathbb{E}(\partial_{\TT_i}\partial_{\TT_j} f )  (\THV_i-{\TTV_0}_i)(\THV_j-{\TTV_0}_j)+ o(|\THV-\TTV_0|^2)= 
		f(\TTV_0) + 1/2 \tr GV + o(|\THV-\TTV_0|^2) $, where $G_{ij}:=(\partial_{\TT_i}\partial_{\TT_j} f )$ is the Hessian.}.
	 In general these are bounds of the form
	\begin{equation*}
	\tr VG \geq f(G) \ , \ \ \forall G\geq 0 \ ,
	\end{equation*} 
	where $V$ is the covariance matrix of the estimator $\THV$ and  $f$ is a real scalar function on semidefinite matrices. 
	This family of inequalities defines  a region in $\mathbb{R}^K$ of allowed values for  the vector of variances $(\var(\TT_1),\var(\TT_2),\ldots,\var(\TT_K))$. The boundary of this region is the trade-off surface. We now show how this is obtained by considering specific examples.

	 \subsection{Classical Trade-off Curves: the Quantum Fisher Information Cram\' er--Rao Bound}
	 By \textit{classical} we refer to the situation when the optimal precision values for the different parameters are independent of each other. This is automatically the case in classical parameter estimation where the maximum likelihood method asymptotically achieves the optimal values for all the variances $\var(\TT_i)$ simultaneously~\cite{Cramer}. 
	 
	 Let us begin by plotting the trade-off curve resulting from the SLD-QFI bound~\cref{eq:MultiParamQFI}. As discussed above, this bound can be interpreted as the assertion that for every direction in parameter space, the single parameter bound applies. Therefore we do not expect to be able to extract nontrivial trade-off relations from it. 
	
	The matrix inequalities~\cref{eq:MultiParamCRBound} and~\cref{eq:MultiParamQFI} imply 
	\begin{equation} \label{eq:2x2QFIExpectedCost}
	\tr NVG \geq \tr \QQ^{-1}G \ , \ \ \forall G\geq 0 \ .
	\end{equation} 
	 Consider the case of two parameters and let $G = \twobytwo{t}{ }{ }{1-t}$ for $t\in(0,1)$. This form of cost matrix corresponds to  a fixed total cost of $1$ which is divided between $\TT_1$ and $\TT_2$ with proportion $t/(1-t)$.  Let  $\QQ^{-1} = \twobytwo{u_1}{b}{b}{u_2}$. \Cref{eq:2x2QFIExpectedCost} becomes
	 \begin{equation*}
	N	(tV_{1}+(1-t)V_{2}) \geq tu_1+(1-t)u_2 \  ,
	 \end{equation*}
	 where   $V_{i}$ is the variance of $\TT_i$.
 	This implies that for every value of $t\in[0,1]$ the points in the $(NV_{1},NV_{ 2})$ plane which are not excluded by~\cref{eq:2x2QFIExpectedCost} lie above the line $NV_{2} =  u_2+ (u_1-NV_{1})\frac{t}{1-t}$. All of these lines pass through the point $(u_1,u_2)$ and as $t$ varies between $0$ and $1$ the slope of the line varies between $0$ and $-\infty$. The allowed region (not excluded by any value of $t$) is $\{NV_{1}\geq u_1\}\cap \{NV_{2}\geq u_2\}$. In particular, the bound~\cref{eq:2x2QFIExpectedCost} does not exclude the point $(NV_{1}=u_1, NV_{2}=u_2)$, which corresponds to optimal precision for both $\TT_1$ and $\TT_2$ simultaneously. This classical---or trivial---trade-off bound is plotted in~\cref{fig:GM2paramTradeoff} as the blue dotted curve.
  	 
\subsection{Non-trivial Trade-off Curves: the Gill--Massar Bound}
 	To demonstrate nontrivial trade-off we shall introduce the bound proved by Gill and Massar in Ref.~\cite{GillMassar}. 
 	They showed that for separable measurements on $N$ identical  copies of finite, $d$-dimensional quantum systems the following  holds:
 	 \begin{equation} \label{eq:trHIbound}
 	  	\tr \FF^{\bm{M}} \QQ^{-1} \leq N(d-1) \ .
 	 \end{equation}
	 This  bound implies~\cite{GillMassar}  that for any $G\geq0$
	\begin{equation} \label{eq:GillandMassar}
		\tr NV G \approx \tr  (\FF^{\bm{M}})^{-1} G \geq \frac{1}{d-1}\left( \tr \sqrt{\sqrt{G}\QQ^{-1} \sqrt{G}}\right)^2 \ .
	\end{equation} 
	We will refer to~\cref{eq:GillandMassar} as the Gill--Massar (GM) bound.

	The non-linear dependence of the right hand side of~\cref{eq:GillandMassar} on $G$ results in a non-trivial trade-off curve. Let  $G$ and $\QQ$ be parametrized as before. Using the following expression for the fidelity of $2\times2$ matrices~\cite{Jozsa} which appears in the right hand side of~\cref{eq:GillandMassar} 
	\begin{equation} \label{eq:Fidelity2x2}
	\left( \tr \sqrt{\sqrt{A}B \sqrt{A}}\right)^2 = \tr AB + 2 \sqrt{\det(AB)} \ , 
	\end{equation}
	 we obtain the following family of lines in the $(NV_{1},NV_{2})$ plane:
	\begin{equation} \label{eq:NonTrivTradeoff}
		tNV_{1}+(1-t)NV_{2} 
		=
		 \frac{1}{d-1} \left( tu_1+(1-t)u_2 +2\sqrt{t(1-t)} \sqrt{\det{\QQ}^{-1}} \right)\ .
	\end{equation}
	To obtain a formula for the trade-off curve fix $V_{1}$ and maximize $V_{2}$ with respect to $t$. This results in the following parametrization of the curve in terms of $t\in(0,1)$:
	\begin{equation*}
		\begin{split}
			NV_{1}(t) &= \frac{1}{d-1}\left( u_1+\sqrt{\frac{1-t}{t}} \sqrt{\det{\QQ}^{-1}} \right) \\
			NV_{2}(t) &=	\frac{1}{d-1}\left(  u_2 + \sqrt{\frac{t}{1-t}} \sqrt{\det{\QQ}^{-1}} \right)
		\end{split}	
	\end{equation*}

\begin{figure}[H]	
			\begin{center}
					\includegraphics [width=0.7\linewidth ]{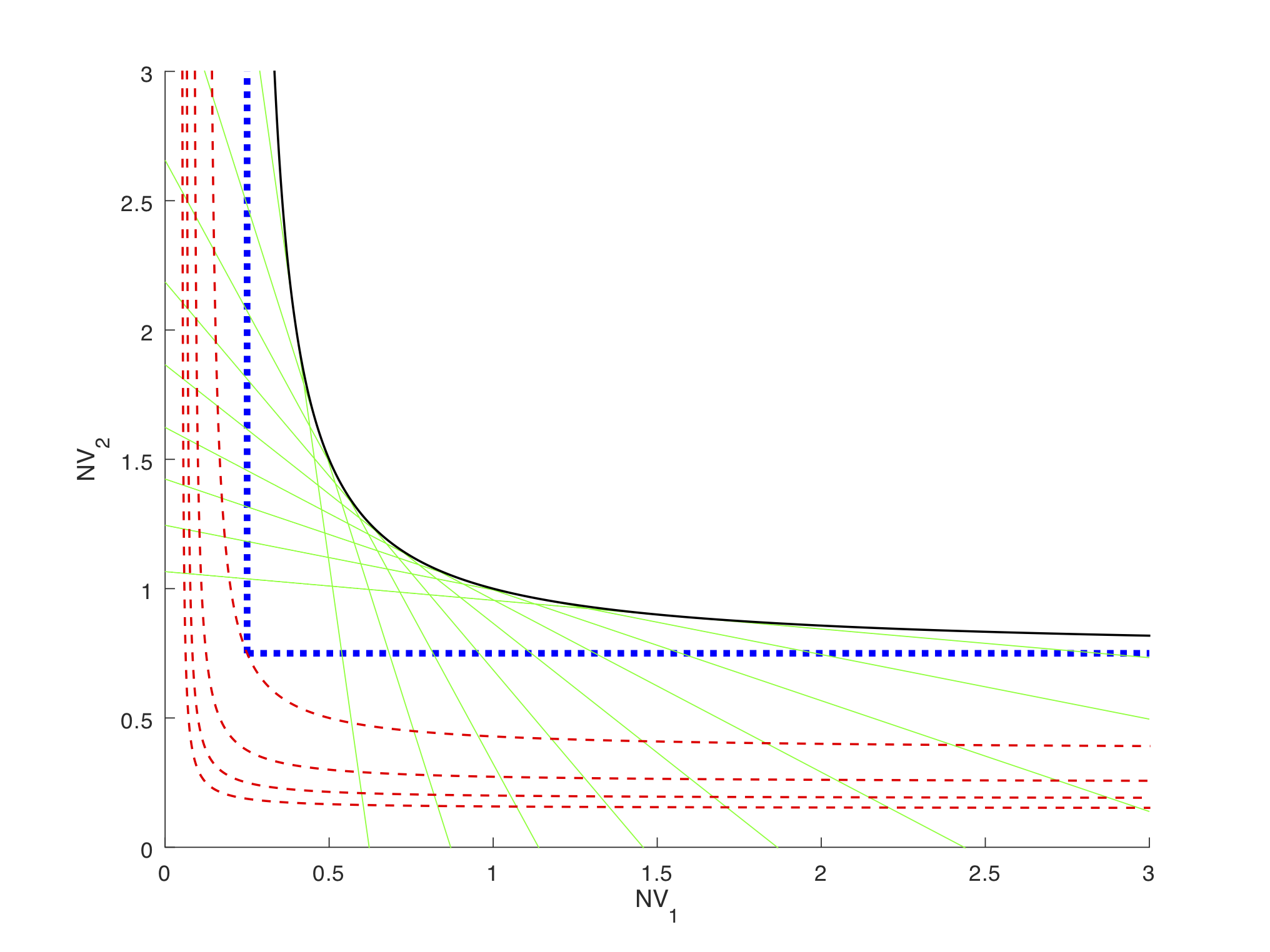} 
			\end{center}
					\caption{Trade-off curves for the rescaled variances of the estimators of a 2-parameter density matrix with Quantum Fisher information matrix $\QQ = \twobytwo{u_1}{b}{b}{u_2}^{-1}$.
						The SLD-QFI bound~\cref{eq:MultiParamQFI} implies the dotted blue   classical trade-off curve.  
						The curves resulting from the GM bound~\cref{eq:GillandMassar} are  plotted for fixed values of $u_1=0.25$ and $u_2=0.75$   for systems of different Hilbert space dimensions: $d=2$ (black solid curve), and  $d=3,4,5,6$ (red dashed curves).
						The GM curves are obtained as the upper envelopes of the lines given by~\cref{eq:NonTrivTradeoff},  several of which  are plotted  (green) for the $d=2$ case.  
				For $d=2$ the     GM trade-off curve  is asymptotic to the SLD-QFI curve, whereas for $d>2$ the GM curves are below the SLD-QFI curve.  }		
			\label{fig:GM2paramTradeoff}
		\end{figure}

		\Cref{fig:GM2paramTradeoff} shows the trade-off curves obtained for fixed values of $u_1,u_2$ and for $d=2,3,\ldots,6$. In addition the trivial trade-off curve resulting from the SLD-QFI bound is shown. The figure clearly shows that for $d>2$ the GM bound is unattainable as it allows a higher precision for each   of the parameters than that allowed by the  SLD-QFI bound. Furthermore,~\cref{fig:GM2paramTradeoff} shows that for $d>2$ the GM bound does not exclude any region above the trivial trade-off curve. This is in agreement whit Ref.~\cite{GillMassar} where  it was concluded  that when the number of  parameters $K$  satisfies  $K\leq d-1$, the SLD-QFI bound is stronger then the GM bound.

			\subsection{The Right Logarithmic Derivative Quantum Fisher Information Bound and the Holevo Cram\' er--Rao Bound}
	\label{sec:TradeOffRLD}
		We shall now introduce  the right logarithmic derivative quantum Fisher information  (RLD-QFI) bound. This bound exhibits nontrivial trade-off, with the 'strength' of the trade-off between the variances of $\TT_i$ and  $\TT_j$  depending directly on the expectation value of the commutator of the corresponding SLDs $\tr \rho[\SS_i,\SS_j]$. 
		 
		The right logarithmic derivative (RLD) is  defined implicitly by
		\begin{equation*}
		\frac{\partial \rho}{\partial\TT_i} = \rho \RR \ .
		\end{equation*}
		The RLD-QFI matrix is then defined by
		\begin{equation*}
		\QQR_{ij} = \tr\rho\RR_j\RR_i^\dagger  \ .
		\end{equation*}
		
		Just as the SLD-QFI matrix, the   RLD-QFI matrix bounds the covariance matrix of any locally  unbiased estimator~\cite{HayashiMatsumoto}:
		\begin{equation*}
		V(\THV)\geq \QQR^{-1}/N \ .
		\end{equation*}
		
		This bound implies, as before, a lower bound on the expected cost associated with any positive cost matrix $G>0$, which, due to the fact that $\QQR$ is a Hermitian matrix (whereas $\QQ$ is real and symmetric) takes the form~\cite[Lemma~6.6.1]{Holevo1982}
		\begin{equation} \label{eq:RLDQFIBound}
		\tr NVG \geq 	 	\tr G\Re\left(\QQR^{-1}\right) + \tr \left| \sqrt{G} \Im \left(\QQR^{-1} \right)\sqrt{G}\right| \  ,
		\end{equation}
		where $|\cdot|$ is the absolute value function defined for Hermitian  matrices via their spectral decomposition;  and $\Re$ and $\Im$ refer to the real and imaginary parts of a matrix taken entry-wise.
		The imaginary part results in a  non-trivial trade-off curve. To see this, consider the case of two parameters.
		Because $\QQR^{-1}$ is Hermitian, its imaginary part is anti-symmetric. Let  
		$\QQR^{-1}  = \twobytwo{r_1}{b+ia}{b-ia}{r_2}$, and $G=\twobytwo{t}{}{}{1-t}$. \Cref{eq:RLDQFIBound} becomes
		\begin{equation}  \label{eq:AbsValExample}
		tNV_{11}+(1-t)NV_{22} \geq 	 	
		tr_1 +(1-t)r_2 + 2a\sqrt{t(1-t)}  \  .
		\end{equation}
		The  right hand side  has  the same functional dependence on $t$ as in~\cref{eq:NonTrivTradeoff} with $d=2$. From this we conclude that this bound results in a non-trivial trade-off curve which is asymptotic to the lines $NV_1=r_1$ and $NV_2=r_2$.
		
		In certain cases, it is possible to express the RLD-QFI matrix in terms of the SLDs. In the case of what is called  a $\mathcal{D}$-invariant model\footnote
		{
		The $\mathcal{D}$ operator is defined implicitly by $ \mathcal{D}(X)  \rho_0+ \rho_0\mathcal{D}(X)    =2i[X,\rho_0]$. A model is called $\mathcal{D}$-invariant if the space spanned by the SLDs is invariant under the action of $\mathcal{D}$. For a further classification of statistical models see Ref.~\cite{Suzuki2019}.
		\label{ft:Dinv}}~\cite{HayashiMatsumoto,Holevo1982}
	 the following holds:
		\begin{equation} \label{eq:RLDfromSLD}
		\QQR^{-1} = \QQ^{-1} +\frac{i}{2} \QQ^{-1} D \QQ^{-1} \ , 
		\end{equation} 
		where $D$ is a matrix whose entries are proportional to  the expectation values of the commutators of the SLDs:
		\begin{equation} \label{eq:Dmatrix}
		D_{ij} = i \tr \rho \left[ \SS_i, \SS_j\right] \ .
		\end{equation}
		As $\QQ$ and $D$ are real, the imaginary part of $\QQR^{-1}$ is $\QQ^{-1}D\QQ^{-1}/2$, which together with~\cref{eq:RLDQFIBound} implies 
		\begin{equation}  \label{eq:RLDboundInTermsOfQFI}
		\tr NV G \geq 	 	\tr G \QQ^{-1}  + \frac{1}{2}\tr \left| \sqrt{G}  \QQ^{-1}  D \QQ^{-1} \sqrt{G}\right| \  .
		\end{equation}
		Comparing to~\cref{eq:AbsValExample} we see that in this case $\tr \rho \left[ \SS_i, \SS_j\right]$ determines how much area the trade-off curve excludes above the trivial curve resulting from the SLD-QFI bound~\cref{eq:2x2QFIExpectedCost} (which has only the $\tr G \QQ^{-1}$ term).
		
		We mention the Holevo Cram\' er--Rao bound, which is in general stronger than both the SLD-QFI and the RLD-QFI bounds~\cite{HayashiMatsumoto}. 
	 In the $\mathcal{D}$ invariant case the Holevo bound  coincides with the RLD-QFI bound~\cite{HayashiMatsumoto,Holevo1982}.
	 As we will be dealing only with such cases, we shall not present the Holevo Cram\' er--Rao bound here and only mention 
	 results we will need for  our discussion\footnote{I addition we mention that it has been recently shown that the bound in~\cref{eq:RLDboundInTermsOfQFI} is always greater or equal than the Holevo Cram\'er--Rao bound, and that the Holevo Cram\'er--Rao bound is less or equal than  two times the SLD bound~\cref{eq:2x2QFIExpectedCost} \cite{Carollo_2019,albarelli2019upper}}.
	The Holevo Cram\' er--Rao bound was shown to be  equal to the SLD-QFI bound iff   the expectation values of the commutators between all  SLDs vanish~\cite{Ragy}. 
	In Gaussian state shift models where one estimates the  displacement parameters, it has been shown that the Holevo Cram\' er--Rao bound is attainable~\cite{Holevo1982}. 
	The theory of local asymptotic normality maps any quantum estimation problem involving many copies of the same state to a Gaussian shift model~\cite{2009CMaPh.289..597K}. 
	This implies asymptotic attainability of the Holevo Cram\' er--Rao bound with \textit{collective} measurements~\cite{Ragy,yamagata2013}.

		\section{The Qubit model}\
		\label{sec:2levelSys}
		Let us next move to the estimation of the most general density matrix of a qubit, which is parametrized by three parameters. This problem is also known as quantum state tomography \cite{MAURODARIANO2003205}. In order to observe trade-off relations between more than two parameters, it is enough to consider a qubit system.  
		In the qubit case, the GM bound is attainable with a measurement performed on single copies of the state~\cite{GillMassar, Hou2016}.

			In this section we compare the GM bound and the Holevo Cram\' er--Rao bound   (which in this case is equal to the RLD bound) through the resulting trade-off surfaces. We also investigate the set of optimal measurements which saturate the inequalities. We characterize this set in two cases: when the parametrization is aligned with $\rho_0$ (when the $z$ axis   is pointing in the direction of the Bloch vector of $\rho_0$); and when it is not aligned. 
		 Finally we use the trade-off surfaces computed for  different parametrizations to obtain a state independent trade-off surface, and derive state independent uncertainty relations.

		We work in the  Bloch sphere  parametrization, using Pauli matrices as a basis, and with $\rho_0 =  [\II+z_0\sigma_z]/2$,  the full parametrization is $\rho(\TT)=\rho_0 +\sum \TT_i \sigma_i$. Note that the initial state can always be brought to this form by rotating the Bloch sphere and working in the appropriate basis. We will call this coordinate system the \textit{adjusted} one, and later---in~\cref{sec:RotatedCoord}---we shall return to describe things in a general coordinate system.
	   We will identify $\TT_1\equiv x$, $\TT_2\equiv y$ and $\TT_3\equiv z$. 
		When the state is full rank ($z_0<1$) the solution to the equation defining the SLDs is unique and given by
			\begin{equation} \label{eq:SLDs}
			\begin{split}
				L_x &=2\sigma_x   \ , 			L_y =2\sigma_y  \\
				L_z &= 2  \twobytwo{(1+z_0)^{-1}}{0}{0}{-(1-z_0)^{-1}}	 \ . 
			\end{split}
			\end{equation}
		The resulting SLD-QFI is diagonal and takes the  form
			\begin{equation} \label{eq:QFIdiag}
		\QQ =4\left(
		\begin{smallmatrix}
		1& & \\
		& 1& \\
		& &  (1-z_0^2)^{-1} 
		\end{smallmatrix}
		\right) \ . 
		\end{equation}

		\subsection{Comparison Between Gill--Massar and Holevo Cram\' er--Rao Bounds}
		Let us take a cost matrix parametrized as 
			\begin{equation} \label{eq:strCostMat}
			G =\left(
					\begin{smallmatrix}
						s & & \\
						  & t& \\
						  & & r:=1-t-s 
					\end{smallmatrix}
			 	\right) \ ; s\geq 0, t\geq 0, s+t\leq 1 \ , 
			 			\end{equation}
		The GM bound is given by~\cref{eq:GillandMassar}:
		\begin{equation} \label{eq:CompGM}
		\tr NVG \geq \frac{1}{4} \left(s+t+r(1-z_0^2) +2\sqrt{ts} +2 \sqrt{r(1-z_0^2)}( \sqrt{t} + \sqrt{s} )\right) \ .
		\end{equation}
		The Holevo Cram\' er--Rao bound is equal to the RLD bound because the model is $\mathcal{D}$-invariant (this is  verified by a direct computation). 
		Computing the matrix  $D_{ij}= i\tr \rho \left[L_i,L_j\right]$ we obtain 
		\begin{equation} \label{eq:DmatrixQubit}
	D=	8z_0 \threebythree{0}{-1}{0}{1}{0}{0}{0}{0}{0} \ .
		\end{equation}
		According to~\cref{eq:RLDboundInTermsOfQFI} the RLD-QFI bound is then given by
		 \begin{equation} \label{eq:CompRLD}
		\tr NVG \geq \frac{1}{4} \left(s+t+r(1-z_0^2) +2z_0\sqrt{ts}\right) 
		\end{equation}
		From  this expression one can already guess that the RLD bound exhibits nontrivial trade-off only between the $x$ and $y$ parameters as $r$ appears only in the  term  coming from $\tr G \QQ^{-1}$ on the right hand side. 
		This is a generic feature of the RLD-QFI bound for finite dimensional quantum systems. We will   show that this is the case in  a $3$-level system in~\cref{sec:3levelSys}.

\begin{figure}[H]	
		\begin{center}
			\includegraphics [width=0.9\linewidth ]{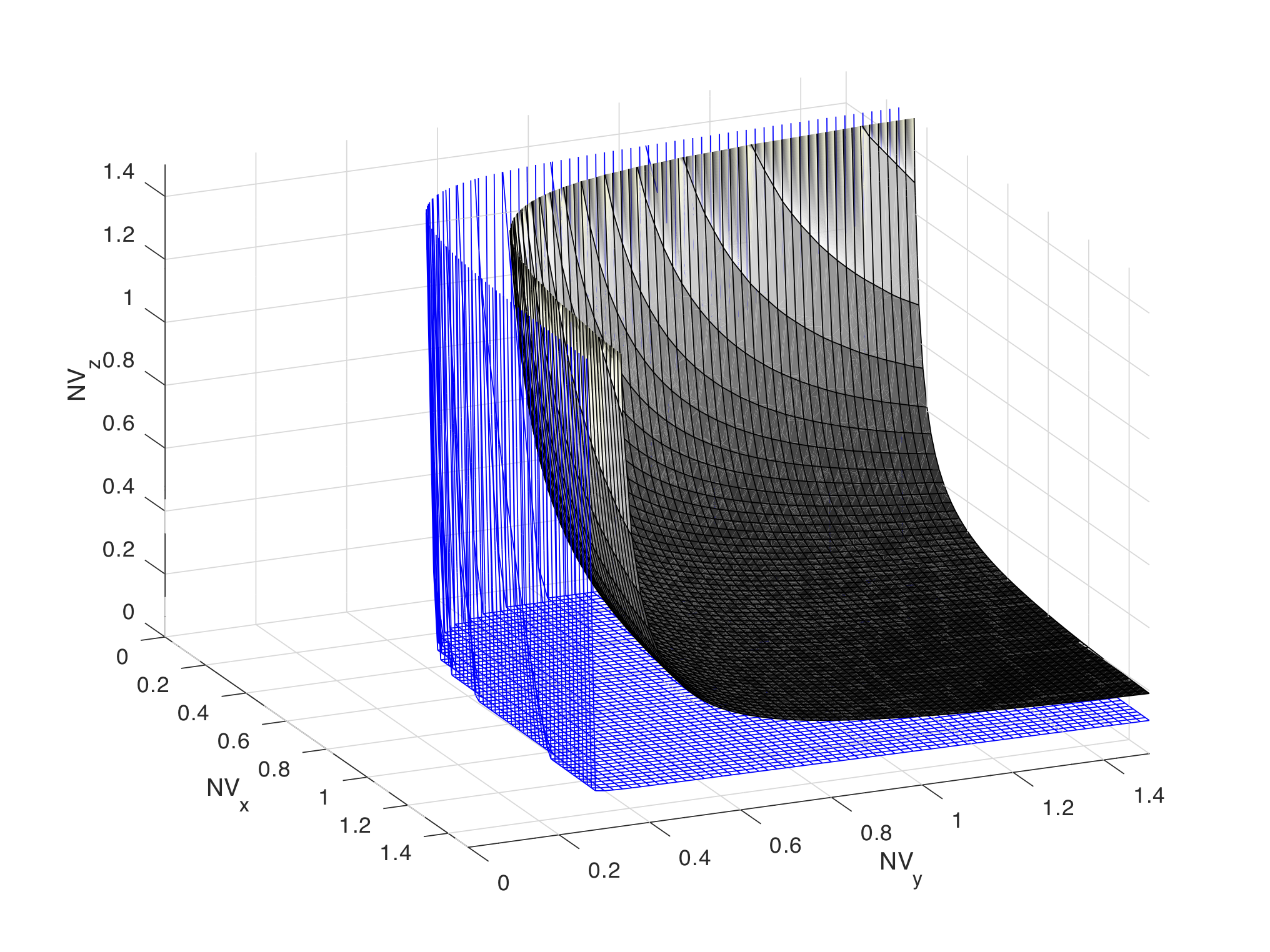} 
		\end{center}
		\caption{ Comparison between the trade-off surfaces obtained from the Gill and Massar (GM) bound (grayscale and filled) and the right logarithmic derivative (RLD) or Holevo Cram\' er--Rao bound (blue and transparent) for estimation of the three Bloch parameters $x,y$ and $z$ for a qubit state   $\rho_0=(\II+0.7\sigma_z)$. The axes  are the rescaled variances of the parameters, i.e.\ the variance multiplied by $N$---the number of copies of $\rho_0$.  The GM trade-off surface  lies strictly above the RLD surface. The RLD surface shows nontrivial trade-off only between the  $x$ and $y$ parameters (it has a 'flat bottom'), whereas the GM bound exhibits nontrivial trade-off between all three parameters simultaneously. Both surfaces are asymptotic to the trivial trade-off surface implied by the SLD-QFI bound~\cref{eq:MultiParamQFI}.
		}	
		\label{fig:GMvsRLD}
	\end{figure}

		Using~\cref{eq:CompGM,eq:CompRLD} 
		we   find for each bound the smallest allowed value of $NV_z$ for a grid of values of $NV_x,NV_y$ (for fixed $NV_x,NV_y$ we can find $NV_z$ by requiring equality in~\cref{eq:CompGM,eq:CompRLD} and maximizing over a grid of values for $s$ and $t$). The results are   plotted in figure \ref{fig:GMvsRLD}. For states with  $z_0<1$,   the Holevo Cram\' er--Rao ($=$RLD) bound is strictly weaker than the GM bound.
		Recall that the GM bound  is attainable with single copy measurements whereas the Holevo Cram\' er--Rao bound with collective measurements. 
		This conforms with our expectation that collective entangled measurements should provide an advantage over separable ones.
		 As the state $\rho_0$ tends towards a pure state, the GM bound tends towards the Holevo Cram\' er--Rao bound, as can be seen from~\cref{eq:CompRLD,eq:CompGM} by setting $z_0=1$.

	\subsection{Measurements Attaining the Gill--Massar Bound}
		The bound~\cref{eq:trHIbound} is achievable for qubits. Gill and Massar show that for a qubit system (d=2), every matrix $\FF$ that satisfies~\cref{eq:trHIbound} is obtainable as the Fisher information matrix of a measurement $\bm{M}^\FF$. $\bm{M}^\FF$ is a probabilistic mixture of three projective measurements along the directions which diagonalize $\FF$ (seen as Bloch vectors). 
		 By probabilistic mixture we mean combining measurements in the following way: let $\bm{M}^{(1)}$ and $\bm{M}^{(2)}$ be measurements with POVM elements $\{M_i^{(1)}\}_{i=1}^I$ and $\{M_j^{(2)}\}_{j=1}^J$. We say that $\bm{M}$ is a probabilistic mixture of $\bm{M}^{(1)}$ and $\bm{M}^{(2)}$  if $\bm{M}$ has $I+J$ POVM elements $M_k = \lambda M^{(1)}_k$ for $k =1,\ldots, I$ and 
				$M_k = (1-\lambda) M^{(2)}_{k-I}$ for $k =I+1,\ldots, I+J$ 
				 for some $\lambda\in (0,1)$; and denote $\bm{M} = \lambda\bm{M}^1\cup (1-\lambda)\bm{M}^{(2)}$.
				 With the obvious generalization to mixtures of more than two measurements.
				This corresponds to measuring $M^{(1)}$ in $\lambda N$ copies of $\rho$ out of an ensemble of $N$ copies, and $M^{(2)}$ on the rest.
				From~\cref{eq:QFisherInfoDef} it is easily verified that the probabilistic mixtures of measurements result in convex combinations of the Fisher information matrices   with the same mixing coefficients, i.e.\ $\FF^{\bm{M}}=\lambda\FF^{\bm{M}^{(1)}}+(1-\lambda)\FF^{\bm{M}^{(2)}}$.

		In the rest of this section, we will require more detailed notation. We denote  the Fisher information matrix corresponding to   $\rho_{\bm{n}} =(\II+\bm{n}\cdot\sigma)/2$, the estimation of parameters $\TTV$, and
		to  a projective measurement $\bm{M}=\bm{P}_{\bm{v}}$ along a Bloch vector $\bm{v}$ as $\FF(\rho_{\bm{n}}, \TTV, \bm{P}_{\bm{v}})$. The following calculation shows that this matrix equals 
		$ \frac{4 }{1-(\bm{n}\cdot\bm{v})^2}\bm{v}   \bm{v}^\intercal$.
		\begin{equation} \label{eq:FisherOfProj}
		\begin{split}
			\FF(\rho_{\bm{n}}, \TTV, \bm{P}_{\bm{v}})_{ij} &:= 
			\frac{\tr P^+_{\bm{v}} \frac{d \rho}{d\TT_i} \tr P^+_{\bm{v}} \frac{d \rho}{d\TT_j} }{\tr P^+_{\bm{v}}\rho} 
			+
		  	\frac{\tr P^-_{\bm{v}} \frac{d \rho}{d\TT_i} \tr P^-_{\bm{v}} \frac{d \rho}{d\TT_j} }{\tr P^-_{\bm{v}}\rho}  = \\
		  	&
		  	2 {v}_i   {v}_j 
		  	\left(
		  	\frac{1}{1+\bm{n}\cdot\bm{v}} +\frac{ 1}{1-\bm{n}\cdot\bm{v}}
		  	\right) = 
		  	4 \frac{{v}_i   {v}_j }{1-(\bm{n}\cdot\bm{v})^2}
		\end{split}
		\end{equation}	
		
			In Ref.~\cite{GillMassar} it is   shown, as part of the proof of the bound~\cref{eq:trHIbound}, that the optimal rescaled covariance matrix for a given cost matrix $G$ is given by
		\begin{equation} \label{eq:OptCov}
			NV_{opt}(G)= \frac{1}{d-1} 
			\left(\tr \sqrt{G^{1/2} \QQ^{-1}  G^{1/2}} \right)
			G^{-1/2}
			 \sqrt{G^{1/2} \QQ^{-1}  G^{1/2}}
			 G^{-1/2}
		\end{equation}
		Plugging in $d=2$,  $\QQ$ from~\cref{eq:QFIdiag} and a cost matrix parametrized as in~\cref{eq:strCostMat} we obtain
		\begin{equation*}  
		NV_{opt}(s,t)= \frac{1}{4} \left(\sqrt{s}+\sqrt{t}+\sqrt{(1-s-t)(1-z_0^2)} \right)
		\threebythree{s^{-1/2}}{0}{0}
								 {0}{t^{-1/2}}{0}
								 {0}{0}{\sqrt{\frac{1-z_0^2}{(1-s-t)}}}
		\end{equation*}
		Comparing this with the Fisher information of a probabilistic mixture with proportions $(\alpha,\beta, 1-\alpha-\beta)$ of projective measurement in the $\hat{\bm{x}}$,$\hat{\bm{y}}$ and $\hat{\bm{z}}$ directions ($\bm{M}(\alpha,\beta) = \alpha P_{\hat{\bm{x}}}\cup \beta P_{\hat{\bm{y}}}\cup (1-\alpha-\beta)P_{\hat{\bm{z}}}$):
		 \begin{equation*}
				\FF(\rho_{z_0 \hat{\bm{z}}}, \TTV, \bm{M}(\alpha,\beta) ) = 
				4 \threebythree
				{\alpha}{0}{0}
				{0}{\beta}{0}
				{0}{0}{\frac{1-\alpha-\beta}{1-z_0^2}} \ , 
		 \end{equation*}
		 we can find $\bar{\alpha}(s,t)$ and $\bar{\beta}(s,t)$ such that 
		 \begin{equation*}
			{\FF(\rho_{z_0 \hat{\bm{z}}}, \TTV, \bm{M}(\bar{\alpha},\bar{\beta})) 
			}^{-1}
		 	=  N	V_{opt}(s,t) \ . 
		 \end{equation*}
		Those are given by
		\begin{equation*}
			\bar{\alpha}(s,t) =\frac{\sqrt{s}}{\sqrt{s}+\sqrt{t} +\sqrt{(1-s-t)(1-z_0^2)}} \  \ ; \   \bar{\beta}(s,t) =\frac{\sqrt{t}}{\sqrt{s}+\sqrt{t} +\sqrt{(1-s-t)(1-z_0^2)}} 
		\end{equation*}
		
		This gives a simple characterization of the optimal measurements, i.e.\ the measurements for which the obtained variances lie on the trade-off surface. They are probabilistic mixtures of projective measurements in the $\bm{x}$, $\bm{y}$ and $\bm{z}$ directions, with different proportions optimizing for different cost matrices. 
		 These projective  measurements happen to be the optimal ones in the  one-parameter  estimation scenario as the SLDs  are diagonal in the $\bm{x}$,$\bm{y}$ and $\bm{z}$ bases respectively (see~\cref{eq:SLDs}). 
		  Note, however, that we have thus far been working in the adjusted coordinate system, where the $z$ axis is aligned with $\rho_0$. 
		  In the next paragraph we 
		  analyze the case of a general coordinate system.

			\subsection{General Coordinates}
			\label{sec:RotatedCoord}
		So far we have considered a general   state $\rho_0$ but in order to simplify the analysis, we adjusted our  coordinate system such that the   $z$ axis was aligned with the Bloch vector of $\rho_0$. 
		 In the last paragraph we saw that in this adjusted coordinate system the optimal trade-off is 
		 attained by probabilistic mixtures of the  Pauli operators (rotated to the adjusted basis). 
		We are not always free to choose the coordinate system we work in, and it is likely that we would like to optimize our  measurement for a cost matrix which is diagonal in a different coordinate system than the one adjusted to $\rho_0$. 
		 We now  look at  the trade-off surface in a coordinate system which is not aligned with the state, and investigate     the  measurements which achieve  the trade-off surface. This will turn out to be useful for deriving our state independent result in~\cref{sec:StateIndep}.
		
		Changing the coordinate system rotates the covariance matrix, the SLDs, and the quantum Fisher information matrix   as described in~\cref{sec:perlimB}. 
		In Ref.~\cite{GillMassar} it is  described how to find the measurement which achieves a desired Fisher information matrix satisfying~\cref{eq:trHIbound} (this is achieved by mixing the projective measurements corresponding to the Bloch vectors which constitute the eigenbasis of the desired Fisher information matrix). We used   their method to compute the optimal measurements  by finding the measurements which result in the inverse of~\cref{eq:OptCov} for different diagonal costs $G$.
		The result is that the optimal measurements no longer belong to an easily characterizable family. As $G$ is varied, the three Bloch vectors describing the projectors of which the measurement is composed travel around the Bloch sphere.
		The SLD measurements, which when working in the adjusted coordinates could be mixed in different proportions to get variances anywhere on the trade-off surface, no longer play a role. Below we demonstrate that they are far from optimal even in the case of a pure cost matrix (one which assigns all the cost  to one parameter), and that in fact, the optimal   for such a cost matrix is to  measure the corresponding Pauli operator.

		In the following we fix an arbitrary coordinate system and test the performance of two families of measurements---one consisting of probabilistic mixtures of the SLD measurements, and the other of the Pauli operators in the chosen coordinate system---and see how they fare compared to the  measurements    attaining the GM bound.
		Let $\rho_0 =(\II+z_0\sigma_z)/2$ as before and let $\TTV$ be the adjusted coordinate system as before. Any orthogonal coordinate system is related to the adjusted one by a rotation. Let  $\tilde{\TT}_i=\sum_j R_{ij} \TT_j$ be the coordinates in which we would like to work, where $R\in O(3)$ is a   rotation  matrix.  The state $\rho$ in these coordinates reads
		\begin{equation*}
			\rho(\tilde{\TTV}) = \rho_0 + \bm{\TTV}\cdot\bm{\sigma} = \rho_0 + R^\intercal  \tilde{\TTV }\cdot \bm{\sigma} = 
			\rho_0 +  \bm{\tilde{\TTV}}\cdot R \bm{\sigma} \ .
		\end{equation*}
		We denote the  Pauli matrices  in the chosen coordinates by $\tilde{\sigma_i}:=(R\bm{\sigma})_i = \sum_j R_{ij}\sigma_j$.
		 As explained in~\cref{sec:perlimB}, the quantum Fisher information matrix now takes the form
		\begin{equation} \label{eq:rotatedQFI}
			\QQ = R \QQ_{diag} R^\intercal  \ ,
		\end{equation}
		where  $\QQ_{diag}$ is the QFI matrix in the adjusted  coordinates given in~\cref{eq:QFIdiag}.
			According to~\cref{eq:LinCombSLD}, the SLDs corresponding to this coordinate system $\SS_{\tilde{\TT}_i}$ are given by linear combinations of the SLDs in the adjusted coordinates:
		\begin{equation*}
			\SS_{\tilde{\TT}_i} = \sum_j R_{ij} \SS_{\TT_j} \ , 			
		\end{equation*}
		Each  $\SS_{\tilde{\TT}_i}$ is diagonal  in a basis consisting of two pure states   corresponding to two antipodal points on the Bloch sphere. For a general rotation $R$,  the three bases diagonalizing  $\SS_{\tilde{\TT}_i}, i=1,2,3$ no longer correspond to three mutually orthogonal lines through the center of the Bloch sphere (because of the non zero $\II$ component in $L_z$, see~\cref{eq:SLDs}).  
 
 		We will now show that the   Pauli measurements in the chosen coordinates achieve the optimal expected cost for a pure cost matrix, i.e.\  $G_i:=\bm{e}_i \bm{e}_i^\intercal$ (where $\bm{e_1}=(1,0,0)^\intercal$ etc.). The optimal cost according to~\cref{eq:OptCov} is given by
 		\begin{equation} \label{eq:OptimalVariancRot}
		\begin{split}
			&		
 			\tr N V_{opt}(G_i)G_i =  
			\left[ 			\tr \sqrt{	G_i\QQ^{-1}G_i }
			\right] ^2 = 
			(\QQ^{-1})_{ii}		= 		\\	&
			\frac{1}{4} (R\threebythree{1}{0}{0}{0}{1}{0}{0}{0}{1-z_0^2}R^\intercal )_{ii}		=
			\frac{1}{4} (R_{i1}^2+R_{i2}^2 +(1-z_0^2)R_{i3}^2) = 
			\frac{1}{4} (1 -z_0^2R_{i3}^2) 
		\end{split}	
		\end{equation}
		where  the last equality is due to orthogonality of $R$.
		The Fisher information for a   Pauli measurement is given by
 	\begin{equation*}
		\begin{split}
		\FF(\rho_0,\tilde{\TTV},\bm{P}_{R^\intercal \bm{e}_i}) = &
		\FF(\rho_{z_0\hat{\bm{z}}},R{\TTV},\bm{P}_{R^\intercal \bm{e}_i}) =
		R \FF  (\rho_{z_0\hat{\bm{z}}},{\TTV},\bm{P}_{R^\intercal \bm{e}_i}) R^\intercal=  \\ 
		&
		\frac{4}{1-z_0^2(R^\intercal \bm{e}_i)^2_z} \bm{e}_i \   \bm{e}_i^\intercal  =
		\frac{4}{1-z_0^2R_{i3}^2 }  \bm{e}_i \   \bm{e}_i^\intercal \ , 
		\end{split} 	
\end{equation*}
	where we used~\cref{eq:FisherOfProj} for the calculation of the Fisher information of a projective measurement in the adjusted coordinates ($\TTV$), and the transformation rule for $\FF$ under change of coordinates.
	 	 Taking probabilistic mixtures of the three   Pauli measurements and inverting the resulting Fisher information matrix we obtain the following family of covariance matrices:
 	\begin{equation*}
 		NV(\alpha,\beta)=
 		 {\FF(\rho_0,\tilde{\TTV},\tilde{\bm{M}}(\alpha,\beta))}^{-1}  =
 		 \frac{1}{4} \threebythree
 		{\frac{1-z_0^2R_{13}^2}{\alpha}}	{0}	{0}
 		{0}	{\frac{1-z_0^2R_{23}^2}{\beta}}	{0}
 		{0}	{0}	{\frac{1-z_0^2R_{33}^2}{(1-\alpha-\beta)}} \ ,
 	\end{equation*}
 	where $\tilde{\bm{M}}(\alpha,\beta)$ is a probabilistic mixture with proportions $(\alpha,\beta,1-\alpha-\beta)$ of the measurements in the  Pauli bases corresponding to our chosen coordinates.
 	We see this achieves the optimal cost for pure cost matrices~\cref{eq:OptimalVariancRot}   for $i=1,2,3$ in the limits $\alpha\rightarrow
 	1$, $\beta \rightarrow 1$ and $\alpha,\beta\rightarrow 0$  respectively.
 
\begin{figure}[H]	
 \centering
 	\subfloat[]{
 	\includegraphics [width=0.5\linewidth]{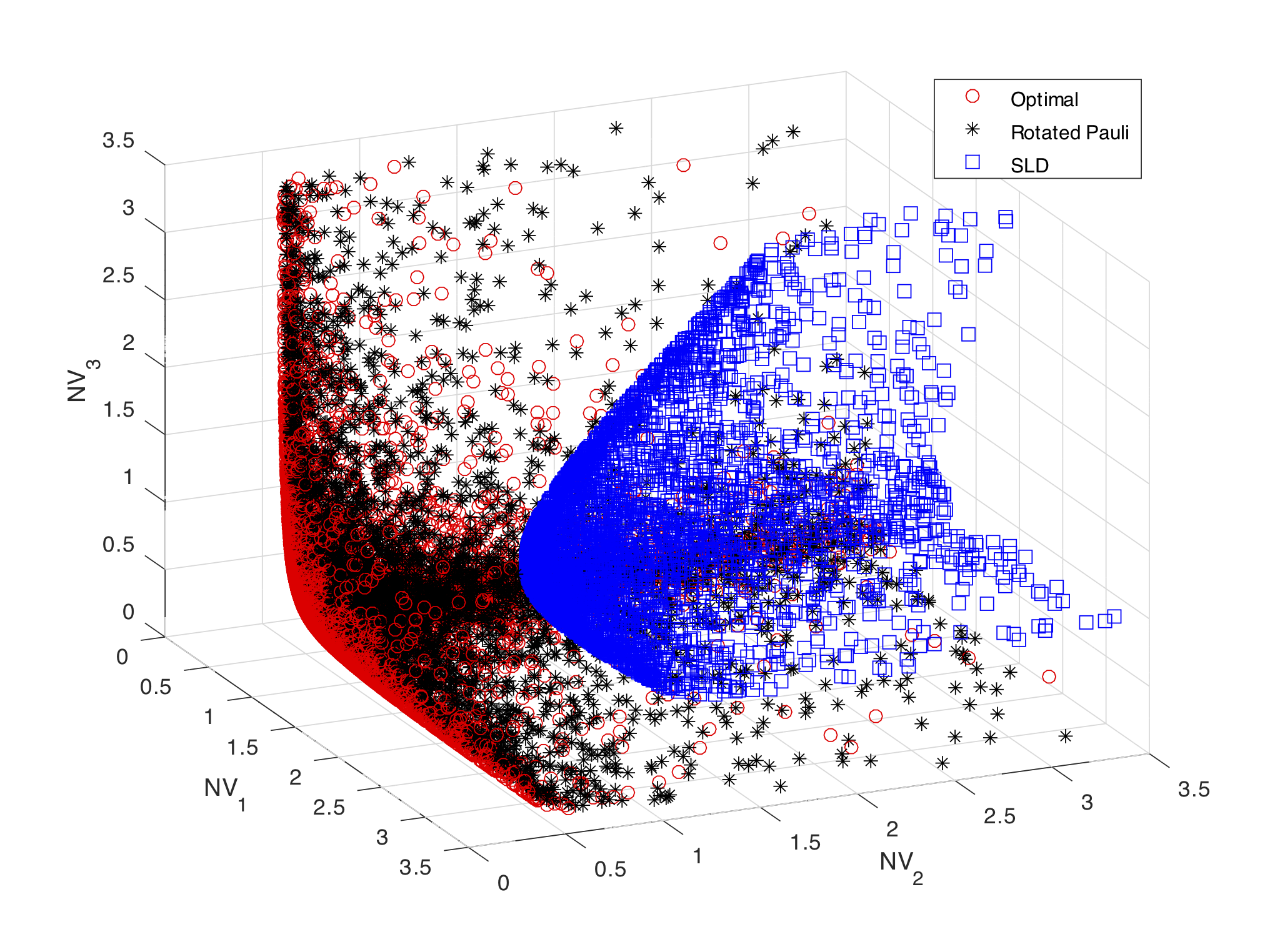} 
 	}
 	\centering
 	\subfloat[]{
 	 	\includegraphics [width=0.5\linewidth ]{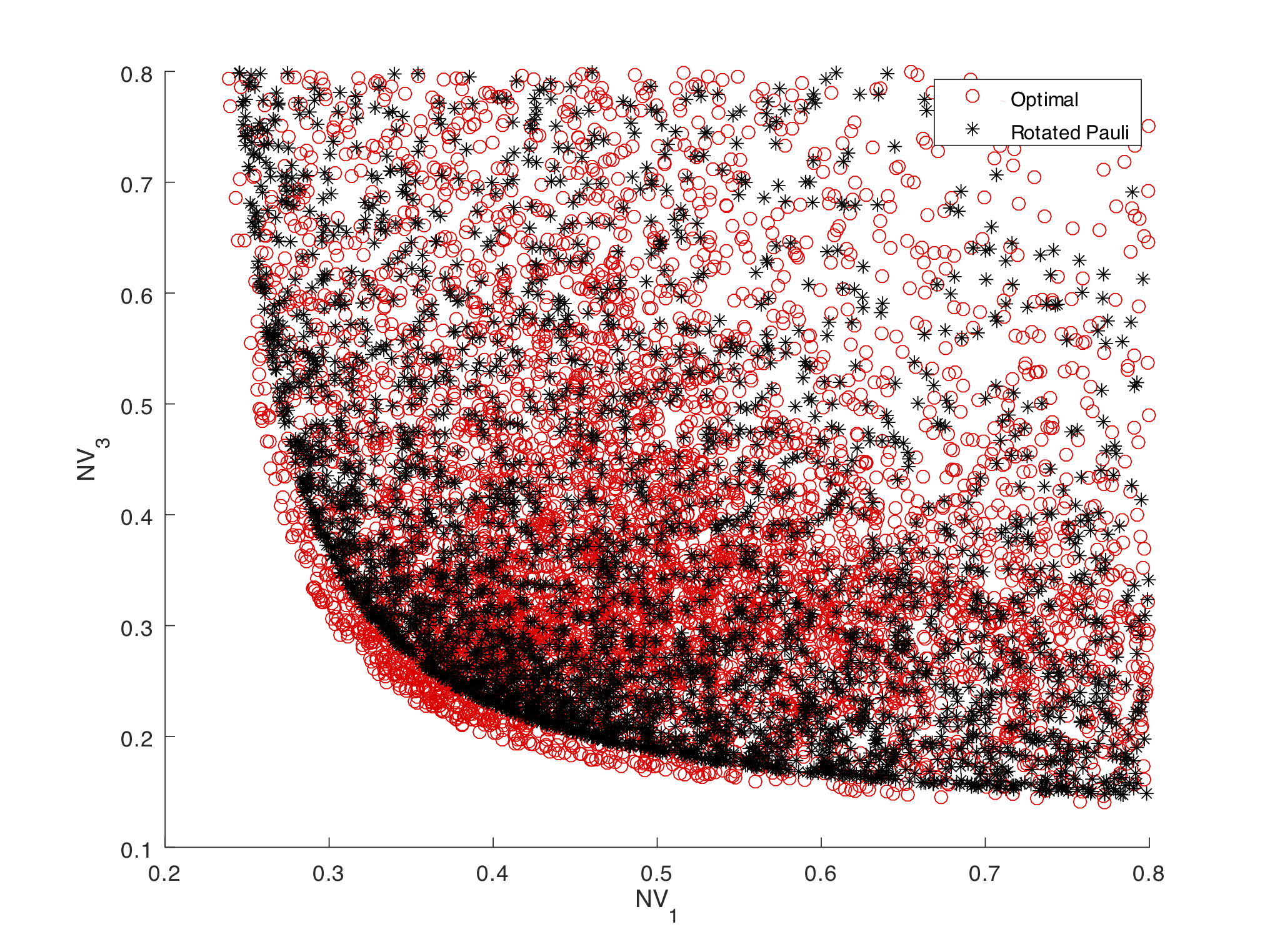}
 	 }
 	\caption{  The variances of the estimators for three parameters of a qubit in a parametrization rotated  with respect to $\rho_0$ resulting from three families of measurements. The variances are obtained as  the diagonal elements of the inverse of the Fisher information matrix $\FF(\mathbf{M})^{-1}$ of a measurement $\mathbf{M}$. 
 		Plot (a) shows the variances of the     measurements optimizing the expected cost for randomly sampled diagonal cost matrices (red circles), these all lie on the   trade-off surface; the variances of  random probabilistic mixtures of rotated Pauli measurements (black asterisks); and random probabilistic mixtures of SLD measurements (blue squares).  Plot (b) is a projection of the points in plot (a) on to the ($NV_1,NV_3$) plane. $\rho_0=(\II+0.92\sigma_z)/2$ and the rotated coordinates  are given  by  $\tilde{\TT}_i=\sum_j R_{ij} \TT_j$, where  $\TT_i$ are the coefficients of the Pauli matrices in the coordinate system aligned with $\rho_0$, and  the rotation matrix $R$ is defined by 		 three Euler angles $R=R_x(\alpha)R_y(\beta)R_z(\gamma); \ \alpha=25^\circ, \beta=25^\circ, \gamma=55^\circ$. It is clearly seen that the SLD measurements perform much worse than the rest, and that the variances of the rotated Pauli measurements lie above, but close to, the trade-off surface. The rotated Pauli measurements approach the trade-off surface far away from the origin as shown in the main text.  }		
 	\label{fig:RotatedCoordinates}
 \end{figure}

 	For randomly sampled  diagonal cost matrices as in~\cref{eq:strCostMat} we computed the optimal covariance matrix using~\cref{eq:OptCov,eq:rotatedQFI,eq:QFIdiag}. In addition we computed the covariance matrices corresponding to random probabilistic mixtures of the Pauli measurements, and to random probabilistic mixtures of SLD measurements. \Cref{fig:RotatedCoordinates} shows the resulting trade-off surfaces between the variances of the three parameters $\tilde{\TT}_i$. It is clearly seen that the   Pauli measurements lie above  the optimal  surface, and that the     SLD measurements preform significantly worse than the other two. This is due to correlations between the parameters in the SLD measurements. 
 	In further numerical calculations we performed it was observed that 
 	the separation between the   Pauli measurements and the optimal measurements is noticeable for  states closer to the sphere of pure states ($z_0 >0.5$), and that   it vanished  when  one of the coordinate axes came close to alignment  with  $\rho_0$.

\subsection{State Independent Trade-Off Surface}
\label{sec:StateIndep}

	So far we have always considered state-dependent bounds. Indeed, all the bounds we used in order to plot our trade-off  surfaces involved explicit dependence on the state $\rho_0$ (recall that the quantum Fisher information matrix $\QQ$ always depends on $\rho_0$). 
	A state independent trade-off surface can be obtained as the boundary of the union over all states $\rho_0$ of the attainable regions---the regions laying above the trade-off surface (equivalently, as the boundary of the intersection of the unattainable regions). 
	To obtain a graphical representation of this state independent trade-off surface, we would need to  plot the trade-off surfaces corresponding to different state $\rho_0$ all on the same plot, and see what region remains uncovered.

	We fix our standard coordinate system to be  in terms of the usual Pauli operators $\rho(\TTV)=\rho_0 +\sum \TT_i \sigma_i$, and for every state $\rho_0$ in the Bloch sphere we use~\cref{eq:OptCov,eq:rotatedQFI,eq:QFIdiag} to sample points from the trade-off surface corresponding to the GM bound with that state. More precisely, we compute the quantum Fisher information matrix $\QQ(z_0)$ in the coordinates aligned with $\rho_0$ ($z_0$ is the length of the Bloch vector of $\rho_0$) and then rotate it back to the standard coordinates with the appropriate  rotation $R\in O(3)$. We  plug the result into~\cref{eq:OptCov} and plot the diagonal entries of $V_{opt}(G)$ for randomly sampled diagonal cost matrices $G$. We do this for a grid of values of $z_0\in[0,1]$ and of the angles parameterizing the rotation ($R=R_x(\alpha)R_y(\beta)R_z(\gamma)$, where $R_x(\alpha)$ is a rotation around the $x$ axis by an angle $\alpha$). This procedure is equivalent to running over a grid of states $\rho_0$.

\begin{figure}[H]	
		\centering
		\subfloat[]{
			\includegraphics[width=0.51\linewidth]{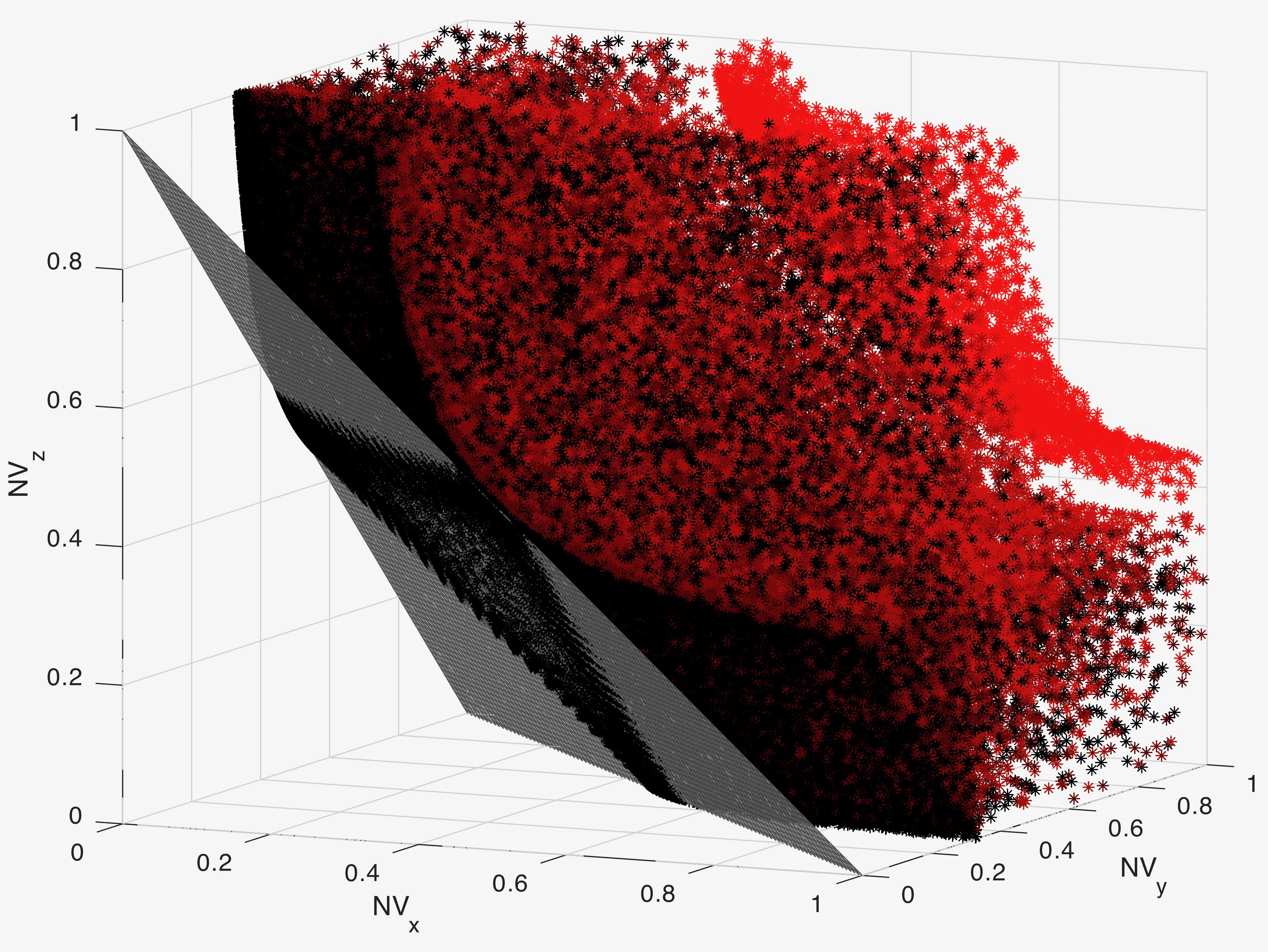} 
		}
		\centering
		\subfloat[]{
			\includegraphics [width=0.49\linewidth ]{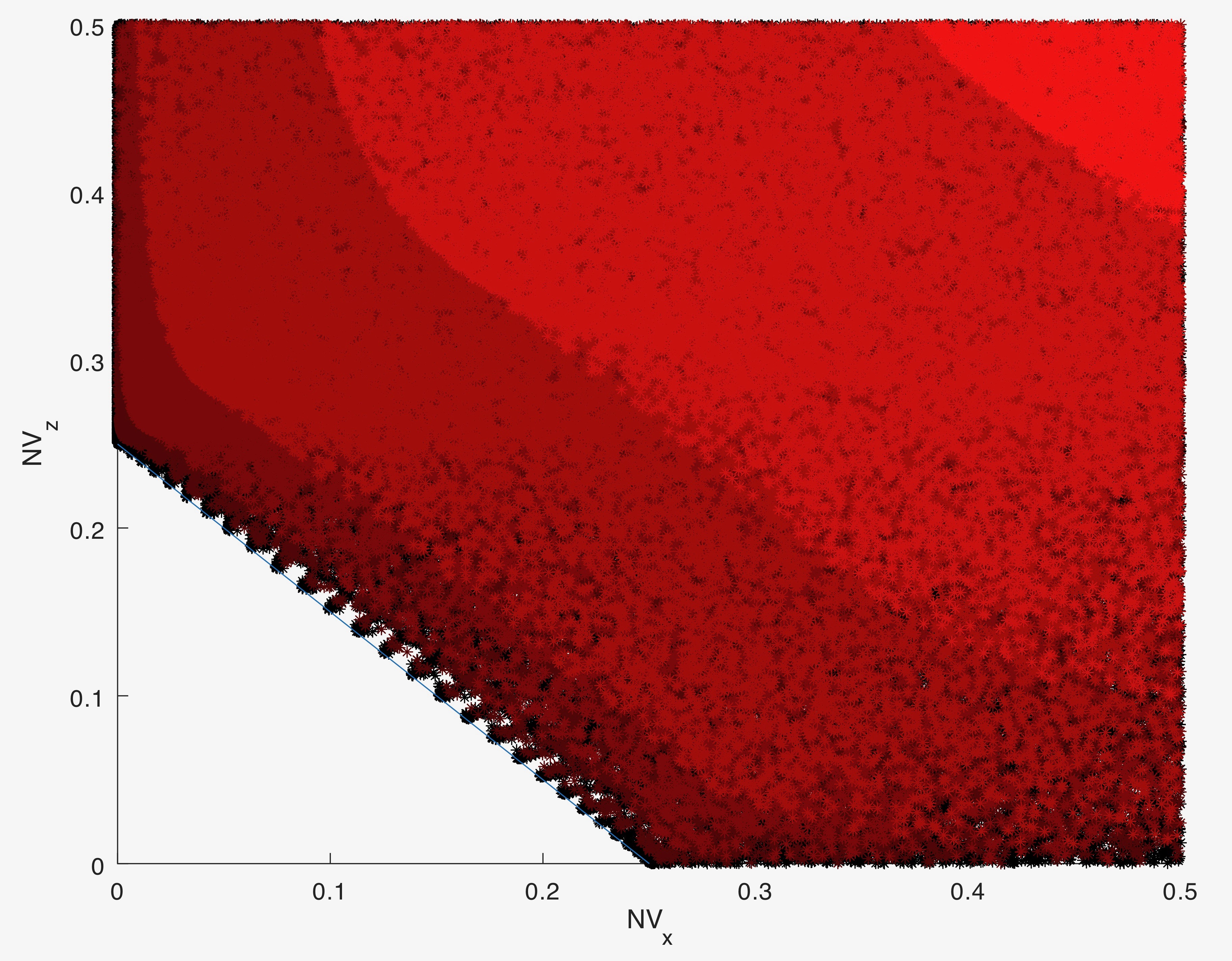}
		}
		\caption{
	 		 State independent trade-off surface. Plot (a) shows points sampled  from      trade-off surfaces corresponding to different states.  The  region which is filled with points is the attainable region, and the region which is empty   is unattainable for all measurements and all states.  
		     The boundary between the regions is the \textit{state independent trade-off surface}. 
		 Plot (a) shows in addition the plane $NV_x+NV_y+NV_z=1$ which forms part of the trade-off surface, as proven in the main text.
		  Black corresponds to purer states and red to states closer to the center of the Bloch ball. 
		Plot (b) is a projection of plot (a) to the $x,z$ plane (black points  were plotted under the  red ones) and shows in addition a straight line  $NV_z=0.25-NV_x$ fitted to the boundary of the region by  maximizing the point of intercept.
			The plots are obtained by random sampling of points from the trade-off surfaces in different coordinate systems, specified by a rotation matrix $R\in O(3)$ which relates the $z$ axes to the Bloch vector of the state $\rho_0$. As explained in the main text, this is equivalent to varying over all states. The values used for the length of the Bloch vector of $\rho_0$ are  $z_0\in \{0.5,0.9,0.99,0.999,1-10^{-4},1-10^{-6}\}$  and the rotations run over a grid 
			of Euler angles:
			$R=R_x(\alpha)R_y(\beta)R_z(\gamma); \ \alpha , \beta , \gamma\in \{0,3^\circ,6^\circ\ldots,360^\circ\}$. The spikes visible on the edge of the covered region in plot (b) are the result of this discrete grid of rotation angles. Pure states ($z_0=1$) were avoided for numerical stability.
		}
		\label{fig:StateIndependentTradeOff}
	\end{figure}

	The result is shown in~\cref{fig:StateIndependentTradeOff}.  The figure shows that a non-trivial state independent trade-off relation holds between the three parameters of a qubit state. This result relies on the GM bound~\cref{eq:GillandMassar} and therefore applies  whenever 	the parameters  are estimated from the outcomes of separable measurement strategies.

	The shape of the  trade-off surface in~\cref{fig:StateIndependentTradeOff}(a) has features similar  to the boundary of the \textit{preparation uncertainty regions} found in \cite[Figures 6,7]{Dammeier_2015}. Its projection to the $x,z$ plane shown in~\cref{fig:StateIndependentTradeOff}(b) suggests that the following uncertainty relation holds for the rescaled variances\footnote{\label{fn:ParameterScaling} Recall that in our parametrization $\TT_i $ is the deviation of $\langle\sigma_i/2\rangle$ from its true value, if we were to parametrize the state as $\rho(\TTV)=\rho_0 + \TTV\cdot\bm{\sigma}/2$ the lower bounds in~\cref{eq:StateIndep2paramBound,eq:StateIndep3paramBound,eq:StatteIndepHolevo} would be $4$ times bigger.}:
	\begin{equation} \label{eq:StateIndep2paramBound}
	NV(\TH_i) + NV(\TH_j) \geq \frac{1}{4}, \;\; i\neq j\in\{x,y,z\} \ .
	\end{equation}
	This bound coincides with the \textit{preparation} uncertainty relation $\Delta(\sigma_x/2)^2+\Delta(\sigma_z/2)^2\geq {1}/{4}$  proven in \cite{BuschPRA2014,Dammeier_2015}.

	We  now prove~\cref{eq:StateIndep2paramBound}.
	It is enough to prove the case $i=1,j=2$, this will become clear from~\cref{eq:VoptDependingOnRandZ0} below, where we have the freedom to rotate $\QQ^{-1}$. We therefore prove $V_1 +V_2 = \tr V P_2\geq 1/4$, where $P_2$ is the following matrix:
		\begin{equation*}
			P_2=\threebythree{1}{}{}{}{1}{}{}{}{0} \; .
		\end{equation*}
		Denote the optimal covariance matrix for a state $\rho_0$ with Bloch vector of length $z_0$ and the cost matrix $P_2$  in a coordinate frame rotated by a rotation $R\in O(3)$  with respect to $\rho_0$  as $V_{opt}( P_2,R,z_0)$. 
		As explained above,  minimizing  the expected cost $\tr V P_2$ over all states $\rho_0$ is equivalent to  minimizing 	$\tr V_{opt}( P_2,R,z_0) P_2 $ over  all choices of coordinate systems (specified by $R\in O(3)$) and all $z_0\in [0,1]$.
		According to~\cref{eq:OptCov} and~\cref{eq:rotatedQFI}  we have
		\begin{equation} \label{eq:VoptDependingOnRandZ0}
		\tr NV_{opt}( P_2,R,z_0) P_2 = \left( \tr \sqrt{\sqrt{P_2}R\QQ^{-1}(z_0)R^\intercal \sqrt{P_2}}\right)^2 \; .
		\end{equation}
	We now proceed to minimize~\cref{eq:VoptDependingOnRandZ0} over $R\in O(3)$ and $z_0\in [0,1]$.	First notice that the minimum is always obtained for pure states ($z_0=1$) because:
	\begin{equation*}  
		\begin{split}
			&\left( \tr \sqrt{\sqrt{P_2}R\QQ^{-1}(z_0)R^\intercal \sqrt{P_2}}\right)^2 
		  	=  
    		\frac{1}{4} \left( \tr \sqrt{ { P_2R} 
    			\left[ P_2+ \threebythree{0}{ }{ }{ }{0 }{ }{ }{ }{1-z_0^2}   			\right]
    			{ R^\intercal P_2} }\right)^2 
    		\geq \\
    		&
    		\frac{1}{4} \left( \tr \sqrt{ {  P_2 R} 
    			  P_2   		 
    			{ R^\intercal P_2} }\right)^2 
    		=
    	\tr NV_{opt}( P_2,R,z_0=1) P_2 \; ,
	\end{split}  
	\end{equation*}
	where we   used the operator monotonicity of the square root function 
	($A\geq B\geq0 \Rightarrow \sqrt{A}\geq\sqrt{B}$) going to the second line.
	We would now like to perform the minimization over $R\in O(3)$. A convenient parametrization of $R$ for this purpose is given by $R(\vec{u},\phi)$ where $\vec{u}$  is a unit vector and $\phi$ is the angle of rotation. Using the Rodrigues' rotation formula  \cite{Goldstein},  $R(\vec{u},\phi)$  is given explicitly by 
	\begin{equation*}
		R(\vec{u},\phi) = \threebythree
		{ c+u_x^2(1-c) } {u_xu_y(1-c)-u_z s} {\ast}
		{u_xu_y(1-c)+u_z s} {c +u_y^2(1-c) } {\ast}
		{\ast} {\ast} {\ast}  \ , 
	\end{equation*}
		where we used the shorthand  $c:=\cos{\phi}$ and $s:=\sin{\phi}$ and where   $\ast$  stands in place of entries we will not use. Plugging this into~\cref{eq:VoptDependingOnRandZ0} and setting $z_0=1$ we obtain 
	\begin{equation*}
	\begin{split}
		  & \tr  N V_{opt}( P_2,R(\vec{u},\phi),z_0=1) P_2 
		   =
		  \frac{1}{4}	\left( \tr \sqrt{ {P_2}R(\vec{u},\phi) P_2 R(\vec{u},\phi)^\intercal  {P_2}}\right)^2 
		=\\
		&
		 \frac{1}{4}	\left( \tr \sqrt{  [R(\vec{u},\phi)]_{12}  [ R(\vec{u},\phi)]_{12}^\intercal   }\right)^2 
			= 
			\frac{1}{4}\tr [R(\vec{u},\phi)]_{12}   [R(\vec{u},\phi)]_{12}^\intercal  + \
			\frac{1}{2}\left| \det [R(\vec{u},\phi)]_{12} \right|  = \\
	&	
	\frac{1}{2}\left[
	 f(c,u_z) + | f(c,u_z)| 
	\right]		 + \frac{1}{4}(1-c^2)(1-u_z^2) ^2 \ ,
	\end{split}  
	\end{equation*}
	where  $[R(\vec{u},\phi)]_{12}$ denotes  the upper left $2\times2$ block of $R(\vec{u},\phi)$, the function $f$ is given by  $f(c,u_z) :=c^2 +c(1-c)(1-u_z^2) + u_z^2(1-c^2) $, and    we used~\cref{eq:Fidelity2x2} to evaluate $\tr\sqrt{\cdot}^2$ for a $2\times 2$ matrix. Our original  minimization problem 
	\begin{equation}
		\min_{|\vec{u}|=1,\phi\in[0,2\pi],z_0\in[0,1]} \left( \tr \sqrt{ {P_2}R(\vec{u},\phi) \QQ^{-1}(z_0) R(\vec{u},\phi)^\intercal  {P_2}}\right)^2  
		\end{equation}
		was therefore reduced to
			\begin{equation}
			\min_{u_z\in[-1,1],c\in[-1,1]} 
			\frac{1}{2}\left[
		f(c,u_z) + | f(c,u_z)| 
		\right]		 + \frac{1}{4}(1-c^2)(1-u_z^2) ^2 \ ,
	\end{equation} 
	which we performed numerically to obtain the value $\frac{1}{4}$.

The two-parameter relations~\cref{eq:StateIndep2paramBound} fully characterize the attainable region for two parameters, as seen from~\cref{fig:StateIndependentTradeOff}(b). As a partial characterization of the shape of the  region attainable for all three parameters in~\cref{fig:StateIndependentTradeOff}(a) we prove the following\footnote{see~\cref{fn:ParameterScaling}}:
\begin{equation} \label{eq:StateIndep3paramBound}
NV(\TH_x) + NV(\TH_y)+ NV(\TH_z) \geq 1 \ ,
\end{equation}
i.e.\ that the plane $NV(\TH_x) + NV(\TH_y)+ NV(\TH_z) = 1 $ is a supporting plane of the attainable region, as can be seen  in~\cref{fig:StateIndependentTradeOff}(a).
As before, the minimum of $\tr V_{opt}( \II,R,z_0) \II$ is obtained when $z_0=1$. There is no need to minimize over $R$ as we can use the cyclicity of the trace to eliminate $R$ with $R^\intercal$. We obtain
\begin{equation*}  
\tr N V_{opt}( \II,R,z_0=1) \II = 
\left( \tr \sqrt{ R\QQ^{-1}(z_0=1)R^\intercal  }\right)^2 =  
\frac{1}{4}\left( \tr \sqrt{ R P_2 R^\intercal  }\right)^2 = 
\frac{1}{4}\left( \tr \sqrt{   P_2   }\right)^2 = 1 \; .
\end{equation*}

We conclude this section by mentioning that the same reasoning can be applied to the Holevo Cram\'er--Rao bound.  Starting from~\cref{eq:RLDboundInTermsOfQFI} with a rotated QFI matrix and a rotated $D$ matrix (\cref{eq:Dmatrix}):
\begin{equation*}
	\tr NV G \geq 	 
		\tr G R\QQ(z_0)^{-1}R^\intercal  + \frac{1}{2}\tr \left| \sqrt{G}  R\QQ(z_0)^{-1} D  \QQ(z_0)^{-1}R^\intercal \sqrt{G}\right| \  ,
\end{equation*}
and setting $G=\II$  we obtain
\begin{equation*}
\tr NV   \geq 	 
\tr    \QQ(z_0)^{-1}   + \frac{1}{2}\tr \left|     \QQ(z_0)^{-1} D  \QQ(z_0)^{-1}  \right| =
  \frac{1}{4} (3-z_0^2 ) +  \frac{1}{2}\frac{8z_0}{16}\tr|\twobytwo{0}{-1}{1}{0}|   = 
  \frac{1}{4} (3-z_0^2 +2z_0 ) \ ,
\end{equation*}
 where we used~\cref{eq:DmatrixQubit}. This is minimized when $z_0=0$ and we obtain the bound\footnote{see~\cref{fn:ParameterScaling}} 
	 \begin{equation}  \label{eq:StatteIndepHolevo}	 	
	 NV(\TH_x) + NV(\TH_y)+ NV(\TH_z) \geq \frac{3}{4} ,
	 \end{equation}
	which is a state independent bound implied by the Holevo Cram\'er--Rao bound and therefore holds for collective measurements.
	It is saturated in the case of a maximally mixed state. In this case the commutation condition $\tr \rho \left[L_i,L_j\right]=0$ is satisfied, which means that the Holevo bound coincides with the SLD-QFI bound \cite{Ragy} and is attainable due to local asymptotic normality \cite{2009CMaPh.289..597K,yamagata2013}.
	This also shows that the state independent trade-off surface  for estimation using collective measurements is different from the one for separable measurements shown in~\cref{fig:StateIndependentTradeOff}, as with collective measurements~\cref{eq:StateIndep3paramBound} can be violated.
	 It would be interesting to compute the state independent trade-off surface implied by the Holevo Cram\'er--Rao bound. We leave this for future works.

	\section{The Holevo Cram\'er--Rao  bound in the qutrit model}
	\label{sec:3levelSys}
	In this section, we compute the Holevo Cram\' er--Rao bound for a qutrit or three level system. We use a model for which the Holevo Cram\' er--Rao bound is equal to the RLD bound and can, therefore, be computed by~\cref{eq:RLDboundInTermsOfQFI}. As in the qubit case, the obtained bound exhibits both trivial and non trivial trade-offs between various parameters. 
 
	We compute the SLDs for a parametrization of a state of a $3$-level system in terms   of the Gell-Mann matrices.
\begin{equation*}
\begin{split}
	\lambda_1 &= \threebythree{0}{1}{0}{1}{0}{0}{0}{0}{0} \; \; \; \; \; 	\lambda_2 = \threebythree{0}{-i}{0}{i}{0}{0}{0}{0}{0} \\
	\lambda_4 &= \threebythree{0}{0}{1}{0}{0}{0}{1}{0}{0} \; \; \; \; \;	\lambda_5 = \threebythree{0}{0}{-i}{0}{0}{0}{i}{0}{0} \\
	\lambda_6 &= \threebythree{0}{0}{0}{0}{0}{1}{0}{1}{0} \; \; \; \; \; \lambda_7 = \threebythree{0}{0}{0}{0}{0}{-i}{0}{i}{0} \\
	\lambda_3 &= \threebythree{1}{0}{0}{0}{-1}{0}{0}{0}{0}  \, \ \	\lambda_8 = \frac{1}{\sqrt{3}}\threebythree{1}{0}{0}{0}{1}{0}{0}{0}{-2}
\end{split}
\end{equation*}
	The state is parametrized as follows:
	\begin{equation*} 
			\rho(\TTV) = \exp\left(-i \sum_{i \in I} \TT_i \lambda_i \right) \rho_d(\TT_3,\TT_8) \exp\left(i \sum_{i \in I} \TT_i \lambda_i \right) \ ,
	\end{equation*}
	where $I = \{1,2,4,5,6,7\}$ contains only indices of non diagonal $\lambda$s and $\rho_d$ is a diagonal state parametrized as\footnote{This parametrization is general enough for local estimation because for a given expansion  $\rho=\rho_0 + \sum t^k\rho_k$, with  $t$ small enough, the following equation 
		$\rho_0 + \sum t^k\rho_k  = \exp(i\sum t^k H_k)(\rho_0 +\sum t^kX_k)\exp(-i\sum t^k H_k) $
		admits a solution with $H_k$ having zero entries on the diagonal and $X_k$ diagonal.}
	\begin{equation*} 
			\rho_d(\TT_3,\TT_8) = \frac{1}{3}\II + (\TT_3+\TT_3^0)\lambda_3 + (\TT_8+\TT_8^0)\lambda_8 \ .
	\end{equation*}
	By choosing $\TT_3^0$ and $\TT_8^0$ we can specify any diagonal   state $\rho_0:=\rho_d(0,0)$. 
	\begin{equation*}
		\rho_0 = \threebythree{k_1}{0}{0}{0}{k_2}{0}{0}{0}{k_3} \ ,
	\end{equation*}
   with $k_3=1-k_1-k_2$. The diagonal entries $k_i$ are related  to $\TT^0_3,\TT^0_8$ by
	 \begin{equation*}
		k_1-k_2 =2\TT^0_3 \ \ , 3(k_1+k_2) -2 = 2\sqrt{3}\TT^0_8 \ .
	 \end{equation*}	
The derivatives of $\rho$, evaluated at $\TTV=0$ are
	\begin{equation} \label{eq:RhoDerivs}
		\begin{split}
			\frac{\partial \rho}{\partial \TT_i} &= -i \left[ \lambda_i, \rho_0  \right] \ \ , i\in \{1,2,4,5,6,7\} \\ 
			\frac{\partial \rho}{\partial \TT_j} &=  \lambda_j \ \ , j\in \{3,8\} 
		\end{split}
	\end{equation}
	In Ref.~\cite{ERCOLESSI20131996} the SLDs for a three level system were computed in a more general setting. For simplicity,  assume the state is full rank, and using the structure constants of $\mathfrak{su}(3)$ (given in Ref.~\cite{ERCOLESSI20131996}), simply verify that      the SLDs are given by
	\begin{equation*}
	\begin{split}
		\SS_1 &= -2\frac{k_1-k_2}{k_1+k_2} \lambda_2 \ \ , 	\SS_2 = 2\frac{k_1-k_2}{k_1+k_2} \lambda_1 \\
		\SS_4 &= -2\frac{k_1-k_3}{k_1+k_3} \lambda_5 \ \ , 	\SS_5 = 2\frac{k_1-k_3}{k_1+k_3} \lambda_4  \\
		\SS_6 &= -2\frac{k_2-k_3}{k_2+k_3} \lambda_7 \ \ , 	\SS_7 = 2\frac{k_2-k_3}{k_2+k_3} \lambda_6 \\
			\SS_3 &= \threebythree{1/k_1}{0}{0}{0}{-1/k_2}{0}{0}{0}{0} \ \ , 
		\SS_8 = \frac{1}{\sqrt{3}}\threebythree{1/k_1}{0}{0}{0}{1/k_2}{0}{0}{0}{-2/k_3} 
	\end{split}
	\end{equation*}
	  When all the $k_i$ are different, the model is $\mathcal{D}$-invariant (see~\cref{ft:Dinv}; this also follows from the $\mathfrak{su}(3)$ structure constants). 
		We compute the expectation values of the commutators of the SLDs $\tr[\SS_i,\SS_j]\rho_0$ to obtain the matrix elements of $D$ (\cref{eq:Dmatrix}). The only non-zero elements are
		\begin{equation*}
		\begin{split}
		|D_{12}| &= 8 \frac{(k_1-k_2)^3}{(k_1+k_2)^2}
		\\
		|D_{45}| &= 8 \frac{(k_1-k_3)^3}{(k_1+k_3)^2}
		\\
		|D_{67}| &= 8 \frac{(k_2-k_3)^3}{(k_2+k_3)^2}
		\end{split}
		\end{equation*}
		The quantum Fisher information matrix  has only two non-zero entries off from the diagonal ($\QQ_{83}=\QQ_{38}\ne 0$). Combining these observations we can use~\cref{eq:RLDboundInTermsOfQFI} to understand the trade-offs which the Holevo Cram\' er--Rao  bound exhibit in this model. We can treat the matrices appearing in~\cref{eq:RLDboundInTermsOfQFI} as block diagonal. In $\QQ^{-1}$ the only block which contains off diagonal terms corresponds to the parameters $\TT_3,\TT_8$. Since in this block, $D$ is zero, we do not need to consider its contribution. In the other blocks (corresponding to $(\TT_1,\TT_2)$,  $(\TT_4,\TT_5)$ and $(\TT_6,\TT_7)$), the result of taking the absolute value of the restriction of  $D$ to this block conjugated with a diagonal positive matrix  (the restriction of $\sqrt{G}\QQ^{-1}$ to the same block), results in a functional dependence of the right hand side of~\cref{eq:RLDboundInTermsOfQFI} which is a sum of three terms similar to~\cref{eq:AbsValExample}. More precisely, for $G_{ij}=\delta_{ij}g_i$ with $g_i\geq 0$ and $\sum g_i =1$
		\begin{equation*}   
		\begin{split}
		\tr GV(\THV) \geq 	& 	\tr G \QQ^{-1}  + \frac{1}{2}\tr \left| \sqrt{G}  \QQ^{-1}  D \QQ^{-1} \sqrt{G}\right|  = \\
		& \tr G \QQ^{-1}  + 
		 \frac{a^2}{2}|D_{12}|\sqrt{g_1g_2} +
		 \frac{b^2}{2}|D_{45}| \sqrt{g_4g_5}  +
 		 \frac{c^2}{2}|D_{67}| \sqrt{g_6g_7}   \  ,
		\end{split}
		\end{equation*}
		where $a,b$ and $c$ are the values of $\QQ^{-1}$ in the corresponding blocks (which we do not compute explicitly as we just want to demonstrate the qualitative behavior). The functional dependence of the above on $\{g_i\}$ implies that non-trivial trade-off appears only between pairs of parameters corresponding to the $x$ and $y$ Pauli matrices within each of the 3 $\mathfrak{su}(2)$ sub-algebras of $\mathfrak{su}(3)$, and within each sub-algebra the trade-off is as the RLD bound in~\cref{fig:GMvsRLD} (trivial trade-off  with the diagonal element).

\section{Discussion}
\label{sec:Discussion}
This paper illustrated the fact that the unsaturability of the   Quantum Fisher information Cram\' er--Rao bound for multiparameter estimation gives rise to a rich variety of quantum uncertainty relations in the form of trade-off curves. 
Those trade-off curves relate to each other the   prefactors $c_{ij}$  of the covariances  $V_{ij}=c_{ij}/N$ of the optimal estimators  for the unknown parameters $\{\TT_i\}$  in the limit when  a large number   of copies $N$ of the  state are available. 
This can be seen as a parameter estimation analogue of the quantum Chernoff and Hoeffding bounds \cite{Audenaert2008} in quantum hypothesis testing, where trade-off curves are obtained for the error exponents $\alpha_i$ for the error of the first versus the second kind ---scaling as $\exp\left(-\alpha_i N \right)$. 

Trade-off curves bring into direct view  the property which distinguishes quantum multiparameter estimation from its classical statistics counterpart---the unattainability of simultaneous optimal precision. 
This property is often discussed in the literature, however, we have never seen such trade-off curves plotted for the known tight bounds.
	Ref~.\cite{demkowiczdobrzanski2020multiparameter} provides a comparison between the Holevo Cram\' er--Rao bound and the Gill--Massar bound by comparing the bounds they put on the expected cost for a single (although state-dependent) cost matrix.  In another work, Ref.~\cite{Lu_PRA_tradeoffs} the authors present the difference between the regions of variances excluded by bounds on their arithmetic, geometric and harmonic means. What distinguishes our approach from the above works is that to obtain the trade-off curve we use the bound on the expected cost for a family of different costs all at once. This is best illustrated in~\cref{fig:GM2paramTradeoff} which shows how the trade-off curve is obtained as the point-wise maximum over a family of lines. We can also apply this in the reverse to obtain  tight bounds on the expected cost given a convex region of attainable variances, as the latter is determined by its supporting hyperplanes.

Trade-offs in quantum parameter estimation  belong to the joint-measurement type of uncertainty relations. They show  that when we wish to estimate certain parameters by performing a measurement on a quantum state, increased precision in one parameter will typically come at the cost of increased uncertainty in other parameters.

Investigation of the trade-off surfaces implied by the Gill--Massar bound led us to our main result---a state independent uncertainty relation between the three parameters of a qubit system. We provided numerical evidence for this trade-off relation (\cref{fig:StateIndependentTradeOff}(a)) and proved an additive bound~\cref{eq:StateIndep3paramBound} which forms part of the trade-off surface. In addition, we proved two-parameter additive uncertainty relations~\cref{eq:StateIndep2paramBound} which coincide with the uncertainty relation for state  \textit{preparation} proven in  Refs.~\cite{BuschPRA2014,Dammeier_2015}. 
 We showed that the Holevo Cram\'er--Rao bound also implies an additive uncertainty relation with a smaller lower bound than in~\cref{eq:StateIndep3paramBound}. Our method for deriving state independent trade-off surfaces from state-dependent bounds could be applied to the  Holevo Cram\'er--Rao bound for a qubit to obtain a trade-off surface for estimation with collective measurements.

The  attainability of the symmetric logarithmic derivative quantum Fisher information (SLD-QFI) bound, which exhibits \textit{classical} (or trivial) trade-off, with collective measurements has recently been shown to be equivalent to the commutation condition $\tr[\SS_i,\SS_j]\rho_0=0$, the vanishing of the expectation values  of the commutators between all SLDs~\cite{Ragy}.
 	The degree to which this fails to be the case has been suggested in Ref.~\cite{Carollo_2019} as a measure of incompatibility between parameters.  
In~\cref{sec:TradeOffRLD} we demonstrated this  by relating  the algebraic form  of the Holevo Cram\' er--Rao bound  to the strength of the resulting trade-off curve.
 \Cref{eq:RLDboundInTermsOfQFI,eq:AbsValExample} show how as the expectation value $\tr[\SS_i,\SS_j]\rho_0$ approaches zero, the corresponding trade-off curve becomes closer and closer to the trivial one. We have also provided  two examples of systems---the qubit (\cref{sec:2levelSys}) and qutrit (\cref{sec:3levelSys}) models---where the commutation condition is satisfied only between some pairs of parameters, and demonstrated how this reflects in their  trade-off surfaces.

The attainable bounds we dealt with in this paper pertain to two different measurement scenarios. The   Gill--Massar bound~\cref{eq:GillandMassar}, is attainable for qubit ensembles ($d=2$) with \textit{separable} measurements, whereas the Holevo Cram\' er--Rao bound~\cref{eq:RLDfromSLD} is attainable for finite dimensional systems with \textit{collective entangled} measurements.
From the algebraic form of the   Gill--Massar bound~\cref{eq:GillandMassar} and the Holevo Cram\' er--Rao  bound~\cref{eq:RLDfromSLD}  the trade-off structure is  not immediately visible. 
In~\cref{fig:GMvsRLD} we   plotted the trade-off surfaces for  each of the bounds for the qubit case to show the qualitative difference between the two. 
The Holevo Cram\' er--Rao bound allows for higher precision and exhibits  non-trivial trade-off only between the $x$ and $y$ parameters, whereas the GM bound---between all three parameters.

The attainability of the Holevo Cram\' er--Rao bound  for finite dimensional systems  relies on the theory of quantum local asymptotic normality. As described  in~\cite{2009CMaPh.289..597K}, in the asymptotic limit the statistical model of a finite dimensional quantum system splits into a product of a classical  Gaussian shift model corresponding to the diagonal elements of the density matrix, and independent harmonic oscillator models for the off-diagonal elements. The   trade-off surfaces of the Holevo Cram\' er--Rao bound, which we described  for the qubit (\cref{fig:GMvsRLD}) and qutrit (\cref{sec:3levelSys}) systems,  are exactly what one would expect to find in the corresponding asymptotic models. 
In both cases the parameters corresponding to the diagonal components  behave like classical systems, i.e. they have trivial trade-off with any other parameter. In the three level system we observe the splitting of the off-diagonal parameters into independent pairs that have non-trivial trade-off within the pair, and trivial trade-off with elements of other pairs.
The trade-off structure described in~\cref{sec:3levelSys} is therefore generic to finite dimensional systems   when  collective measurements can be implemented.

Finally, we studied the optimal single copy measurements in the qubit model. We showed that the   strategy of measuring different SLD operators on parts of an ensemble of identical states, which is optimal for the case of a coordinate system aligned with the state $\rho_0$, is far from optimal in the   case of general  coordinates. 
We  further demonstrated that measuring the Pauli operators   (rotated to the coordinate frame)   achieves the optimal cost when all the cost is assigned to one parameter. Our numerical calculations~\cref{fig:RotatedCoordinates} further  showed  that the rotated Pauli measurements are   not very far from the optimal for general cost matrices.

\section{Acknowledgments}
The authors thank Lorenzo Maccone for the suggestion to look for a state independent bound from  state-dependent trade-off curves,  and Borivoje Dakic for helpful discussions. We   acknowledge the support of the Austrian Science Fund (FWF) through the Doctoral Program CoQuS and the project I-2526-N27 and the research platform TURIS.
I.K.\ acknowledges support from the European Commission via Testing the Large-Scale Limit of Quantum Mechanics (TEQ)  project (No.\ 766900). 
 

%

\end{document}